\documentclass[AMA,STIX1COL]{WileyNJD-v2}
\usepackage{moreverb}
\usepackage{mathtools}
\usepackage{url}
\usepackage{algorithm}
\usepackage{amssymb}
\usepackage{amsmath}
\usepackage{subfig}
\usepackage{makecell}
\usepackage{tikz}

\newcommand{\invres}[5]{
  \draw[|-|, thick] (#2,#4) -- (#3,#4); 
  \node[above right] at (#2,#4) {#1}; 
  \node[left] at (#2,#4) {$inv_#5$}; 
  \node[right] at (#3,#4) {$res_#5$}; 
}

\newcommand{\mathtxt}[1]{
  \textit{#1}
}

\articletype{Article Type}%

\received{<day> <Month>, <year>}
\revised{<day> <Month>, <year>}
\accepted{<day> <Month>, <year>}

\begin{document}

\title{Efficient Linearizability Checking for Actor-based Systems}

\author[1]{Mohammed S. Al-Mahfoudh}

\author[2]{Ryan Stutsman}

\author[3]{Ganesh Gopalakrishnan}

\address[1,2,3]{\orgdiv{School of Computing}, \orgname{University of
    Utah}, \orgaddress{\state{Utah}, \country{United States of
      America}}}

\corres{Mohammed S. Al-Mahfoudh and Ryan Stutsman, \email{\{mahfoudh,stutsman\}@cs.utah.edu}}


\abstract[Summary]{Recent demand for distributed software had led to a surge in popularity in
actor-based frameworks.  However, even with the stylized message passing model
of actors, writing correct distributed software is still difficult.  We present
our work on linearizability checking in DS2, an integrated framework for
specifying, synthesizing, and testing distributed actor systems.  The key
insight of our approach is that often subcomponents of distributed actor systems
represent common algorithms or data structures (e.g.\ a distributed hash table
or tree) that can be validated against a simple sequential model of the system.
This makes it easy for developers to validate their concurrent actor systems
without complex specifications. DS2 automatically explores the concurrent
schedules that system could arrive at, and it compares observed output of the
system to ensure it is equivalent to what the sequential implementation could
have produced.
We describe DS2's linearizability checking and test it on several concurrent replication algorithms from the literature. We explore in detail how different algorithms for enumerating the model schedule space fare in finding bugs in actor systems, and we present our own refinements on algorithms for exploring actor system schedules that we show are effective in finding bugs.
}



\maketitle

Recent years have seen a surge in actor-based programming.
%
This is,
in part, because many applications today are naturally concurrent and
demand some means for distribution and scale-out.
Hence, actor-based
frameworks~\cite{akka,erlang,orleans-paper,actor-model-partisan}
have seen large-scale production
deployment~\cite{akka-fortnite,orleans-halo} running real, low-latency
services for thousands to millions of concurrent users.
However, despite a stylized message-passing-based concurrency scheme
that avoids the complexity of shared memory, actor systems still
contain bugs~\cite{bita}.
As a result, prior works have sought to bring model checking
techniques~\cite{agha:transdpor,basset} and code
coverage-driven techniques~\cite{bita} to bear on these systems with
some success.
%

However, beyond exploring schedules efficiently, a separate challenge
is the more basic problem of specifying invariants and safety
properties for these types of complex systems~\cite{verdi,coq}.
One approach that has gained traction in recent years is manual,
black-box testing of systems under complex, concurrent scenarios and
comparing against a simple, sequential reference implementation that
meets the same interface as the complex, concurrent
system~\cite{jepsen}.
For example, recording and checking the
linearizability~\cite{herlihy:linearizability}
of a \emph{history} of responses produced by the larger system against
responses produced by the reference implementation provides a sound
and complete means for testing \emph{a single, particular execution
schedule of the system}~\cite{pcomp,wg:linearizability}.
The problem with this approach is that if interesting schedules are
missed by the user testing the system or if certain schedules are hard
to produce, then this approach says little about the overall
correctness of the system.


Our goal is to combine the success of these two techniques (schedule
exploration and linearizability checking) in an automated
package for actor systems.
These systems often contain subcomponents that provide
concurrent implementations of abstract data types (ADTs) with
well-defined inputs and outputs (\emph{invocations} and
\emph{responses}).
Hence, these subcomponents are amenable both to
exhaustive schedule generation \emph{and} to automated testing against
reference implementations of the corresponding ADTs.
%
This new approach provides a richer set of checks than simple schedule
exploration against user-specified assertions while maintaining simplicity of user provided specification.
%
Of course, to be practical, such an approach must control explosion both in
schedule exploration and in history checking costs.
%


As a first step toward this goal, this paper attempts to understand how best to
control that explosion by assessing the performance of many classic and
state-of-the-art schedule exploration algorithms when considered together with
state-of-the-art linearizability checking algorithms.
To that end, we compare several algorithms on multiple
actor systems.
For example, we test with a simple distributed register and compare checking costs
when that same ADT is lifted for fault tolerance using standard consensus-based
state machine replication
protocols~\cite{raft,viewstamped.revisited,paxos-simple}. 
This shows that beyond testing individual systems, this approach also works as
an easy approach to testing these notoriously subtle black-box replication
techniques.
%
Ultimately, we show that by 1) subdividing systems by the principle of
compositionality of
linearizability~\cite{lulian:linearizability,pcomp,herlihy:linearizability} and
2) exploiting independence between actors for effective dynamic partial order
reduction (DPOR)~\cite{agha:transdpor,flanagan:dpor}, we can limit the number
of schedules to make finding bugs in concurrent actors systems practical by
comparing against a simple, sequential reference implementation of the system's
ADT.


Beyond this first step, our analysis informs a larger effort on our
framework, an actor-based framework for specifying and checking distributed
protocols.
It provides a concise language for specifying systems, and
using our framework, users can manually drive testing on their systems
(the framework includes some specialized schedulers specifically for
this kind of test-driven checks), can rely on its automated schedule
exploration and linearizabilty checking, or can extend its automated
schedulers.
Our end goal is an easy system for specifying and testing distributed
actor systems that can drive synthesis to practical implementations.


In sum, this paper makes the following contributions:
\begin{itemize}
\item Our framework~\cite{ds2:repo} that not only made our schedulers fully
stateful\footnote{Saving the entire state of the distributed system in order to
back track to it during exploration, instead of the previous implementations of
algorithms that re-run the system controllably from start to end.},
extensible\footnote{Both hierarchies of schedulers/algorithms and distributed
system model are extensible independently and they will still be compatible with
each other.}, re-usable\footnote{Class hierarchies are modularily structured
that one can override few methods while keeping the rest intact and still be
re-usable.}, composable\footnote{Schedulers can be composed, nested, and can
save the state and do hand-off between themselves without having to restart
explorations.}, and modular\footnote{Schedulers are structured in a way that all
their behavior is overridable in parts or in whole without having to re-write
all parts of the scheduler e.g. the main loop.} but also reduced our novel
algorithms/schedulers implementations to mere \emph{predicates} on receives
(i.e. \texttt{didEnable(receive1,receive2)} and
\texttt{areDependent(receive1,receive2)}), one for each of our algorithms. These
features make schedulers and the networked distributed system model (the
context) simple for developers to extend. 
%
\emph{practically}. The novelty in this work does not stop at an
almost no overhead algorithm to dynamically detect causality between events
(zero memory overhead, and almost no compute overhead per the experiments) and
at the same time deduplicating retries. It goes beyond that with a unique
declarative environment for exploring actor systems in Scala. This promotes
formalism in software practice.

\item An automated approach to correctness testing of subcomponents of
  actor systems no developer-provided specification aside from a simple,
  non-concurrent (i.e.\ sequential) specification reference
  implementation of the component and a simple test harness.

\item An exploration of our approach and its effectiveness on several
  practical ADTs including quorum-replicated
  registers~\cite{dynamo,cassandra,giffords-algorithm}, a model of
  Open Chord~\cite{benchmark:openchord:heath,benchmark:openchord:zepeng,make-chord-correct,chord},
  and Multi-Paxos~\cite{chandra:paxos-live,paxos-simple,paxos,benchmark:paxos}.

\item Extensible and modular implementations of 7 algorithms (6 of
  which are \emph{stateful}), and two of which are new (\emph{IRed} and \emph{LiViola}). All
  of which used only four out of sixteen (programmable) operational
  semantics rules~\cite{ds2-opsem} provided by our framework~\cite{ds2-model,ds2:repo,ds2-and-opsem-report}.
  %

\item A detailed comparison of these algorithms used for the first
  time in the context of linearizability checking instead of simple
  invariant checking, run side-by-side on the same benchmarks showing
  how they perform and scale when checking actor systems. This helps
  show effectiveness of focusing schedule generation towards revealing
  linearizability violations.
\end{itemize}


\section{Background}
\label{sec:background}

\subsection{Actor Systems}

Actors are a model for specifying concurrent systems where each \emph{actor}
or \emph{agent} encapsulates some state.
They do not share state and have no internal concurrency\footnote{Internal
concurrency is any concurrency inside a single actor e.g. multithreading. An
actor is strictly a sequential communicating process. Mixing multithreading
breaks the actor model's encapsulation.}.
Instead, actors send and receive \emph{messages} between one another.
Internally, each actor receives an incoming message and performs an \emph
{action} associated with that message.
Each action can affect the actor's local state and can send messages
to other/same actor(s).
Each actor executes actions sequentially on a single thread,
but an actor system can run concurrent and parallel actions since the
set of receiving actors collectively perform actions concurrently.

By eliminating complex constructs like threading and shared memory, actors
make it easier for developers to reason about concurrency.
For example, data races\footnote{We distinguish between a data race
(which is a race condition over the data in the local state of an
actor) and a race condition (which is any non-deterministic behavior
due to the order in which events are processed)} are impossible in actor
systems.
However, race conditions can manifest in the order of the messages
received by the actor in case they interfere.
This programming model naturally supports distribution, since it
relies on message passing rather than shared memory.
As a result, there are many popular actors
frameworks~\cite{akka,orleans-paper,agha:actors,hewitt:actors,actor-model-wiki}
that closely adhere to the actor model.

\subsection{Model Checking} 

Model checking has been applied in many domains to assess the correctness of
programs~\cite{samc,pavlo2011,mc,modist,10.1145/1047659.1040315,}. This includes
actor systems~\cite{agha:transdpor}, real-time actor languages~\cite{rt-actors},
and the Rebeca Modeling Language~\cite{rebecca}.
Model checkers explore states a system can reach by systematically
dictating different interleavings of operations (each of which is called a {\em
schedule}).
In actor systems, concurrency is constrained by the set of
sent-but-unreceived messages in the system, which determines the set
of \emph{enabled actions} at each point in the execution.
Hence, it consists of exploring the set of all possible interleavings
of message receives.

This brings the issue that we are assuming a hand-shake driven model.
This is true, but our model (and its operational semantics rules)
is not limited to that, as it offers \emph{timed actions} (real time,
since it is executable, or relative ordering) for blocking and
non-blocking actions.
However, since that would need users' modulation of input
application times, schedulers do \emph{not} implement this yet.

In model checking, exploration is typically coupled with a set of invariants or
assertions provided by the developer of the system under check.
By checking that the invariants hold in all states reachable via any
schedule, developers can reason about safety and/or correctness
properties of their program.

Many different strategies have been explored for guiding schedule enumeration
or reducing its inherent exponential cost.
Ultimately, for non-trivial programs model checkers cannot enumerate all
schedules and must bound exploration; hence, this may lead to algorithms that
are sound but with incomplete results, i.e. not complete state space coverage.
Randomized scheduling has shown promise, since it explores diverse sets of
schedules.
Another approach is dynamic partial order reduction~(DPOR)~\cite{flanagan:dpor}, which
prunes schedule enumeration by only exploring enabled actions that could
interfere\footnote{We say messages ``interfere'' when they are
received by the same actor and their effects do not commute. Similarly in
threads, two operations interfere when there is a write operation whose
effect(s) does not (do not) commute with another write/read operation.} with one
another.
This has been extended to the context of actor-based systems where the extra
independence between actions (due to the lack of shared memory) allows
additional pruning~\cite{agha:transdpor}.
We describe several of these strategies in more detail in
Section~\ref{sec:overview}.

Importantly, before a developer can use a model checker to check their code for
correctness, they must first specify properties that the scheduler should check
as it explores systems' states.
This is a challenge for most developers, especially in concurrent
systems.



\subsection{Linearizability}

Linearizability is a consistency model for concurrent objects (e.g.\
registers, stacks, queues, hash tables) ~\cite{herlihy:linearizability}.
%
%
Linearizability has several key properties that makes it common and popular in
distributed programming.
It is strict about ordering, which eases reasoning, but it allows enough
concurrency for good performance.

From the perspective of an object user (or a system representing
it), each operation they invoke appears to happen atomically
(instantaneously) between the time of its \emph{invocation} until the
time of its \emph{response}, a \emph{linearization point} in time.
%
%
Because operations take effect atomically, i.e. totally orders them,
the concurrent object has a strong relationship to a sequential
counterpart of the same ADT.
For some \emph{history} of operations (a sequence of invocations and
responses) on that object, there must be a total order of those
operations that when applied to a sequential implementation of the
same ADT produces the same responses.
This is powerful because any sequential implementation 
can be used to cross-check the responses of a concurrent
implementation against its sequential counterpart.


\begin{figure}[t!]
  \centering
  \input{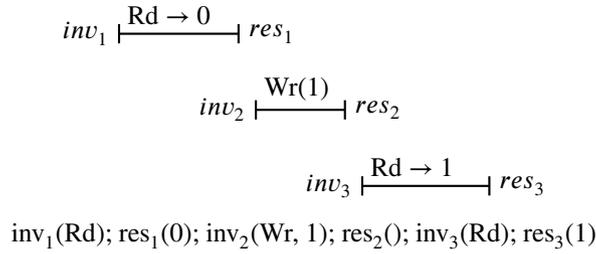}
  \caption{A sequential history of operations on a register. \textbf{inv} stands for invocation. \textbf{res} stands for response. \textbf{Rd} stands for a read operation. \textbf{Wr} stands for a write operation. Numbers appearing on those operations correspond to the order at which the invocation was issued.}
  \label{fig:reg1seq}
\end{figure}

\begin{figure}[t!]
  \centering
  \input{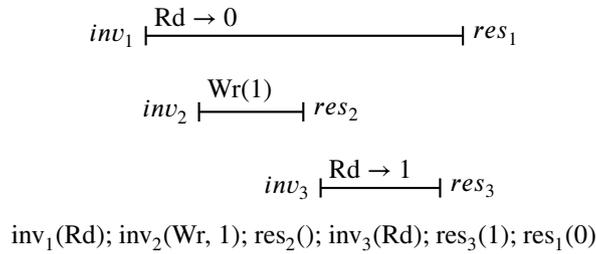}
  \caption{A concurrent history of operations on a register. This history is
    \emph{linearizable}; it produces an equivalent effect as the sequential
    history in Figure~\ref{fig:reg1seq}.}
  \label{fig:reg1conc}
\end{figure}


Figure~\ref{fig:reg1seq} visualizes a \emph{sequential} history of operations
on a register that supports a read and write operation.
Figure~\ref{fig:reg1conc} shows a \emph{concurrent} history using the
same abstract type (a register).
The operations overlap and run concurrently, but the register produces the same
response to each of the invocations as the history in Figure~\ref{fig:reg1seq}.
Hence, the concurrent register executed the operations in a consistent
manner to the sequential counterpart.
The user can reason about concurrent operations on the register in
similar way to that of a sequential implementation protected by a
mutex.
For example, in Figure~\ref{fig:reg1conc} res$_1$ could return 1, in
which case the execution would be equivalent to a different sequential
history; this is okay.
However, res$_3$ could never return 0, since no sequential history
where inv$_3$ happens after the completion of res$_2$ could explain
that result; otherwise, this would indicate a bug in the concurrent
implementation.
Importantly, all of this can be understood by observation only, and
basic understanding of the equivalent sequential reference
implementation (for invocations vs responses).

\textbf{Linearizability Checking and WGL Algorithm~\cite{lowe:wgl:linearizability}.}
This correspondence between sequential and concurrent histories is what
enables automated detection of bugs in concurrent objects.
%
To do this, one can capture a history of operations from a concurrent structure.
Feeding these operations one-at-a-time into a sequential structure,
say, in invocation order may produce the same return values for each
response, in which case the captured history is consistent with
linearizability.
However, this might not work because the concurrent execution
may lead to different responses orderings.
%
Even repeating the same procedure in response order can be fruitless.
%
Only an exhaustive search over the space of potential equivalent
histories may yield a correspondence.
%
WGL algorithm~\cite{lowe:wgl:linearizability} generates these permutations which are all
of the same history that 1) never reorders a response before its
invocation, and 2) never reorders two invocations.
For each sequential history it finds (a history where each invocation
is adjacent to its response) it feeds the history to a sequential
implementation of the ADT being checked to see if all the responses match.
If some sequential history that explains the concurrent one is
discovered, then the implementation being checked behaved in agreement
with linearizability \emph{in the execution described by that one
history.}
Otherwise, the history is deemed non-linearizable.

\section{Overview}
\label{sec:overview}

The focus of this paper is to explore model checking in the
context of linearizability checking.
That is, model checking can be used to systematically produce
histories.
When put together, these techniques would let developers check full,
concurrent actor systems for correctness without manual specification
of invariants.
The key problem is that both algorithms are exponential; however, this
says little about the potential usefulness of combining the
techniques.
Past works have proposed many ways to reduce schedules to explore.
No prior work explores how these different schedulers impact the set
of histories to check when exploring a structure, nor does any prior
work explore the interplay between model checking costs and history
checking costs.
Hence, we begin our efforts to improve linearizability checking costs
for practical actor-based systems with a quantitative exploration of
existing techniques.
Later, we describe our own new schedulers designed to improve over
them.


\subsection{Toolchain Flow}
\begin{figure}[b!]
  \centering
  \includegraphics[width=\linewidth]{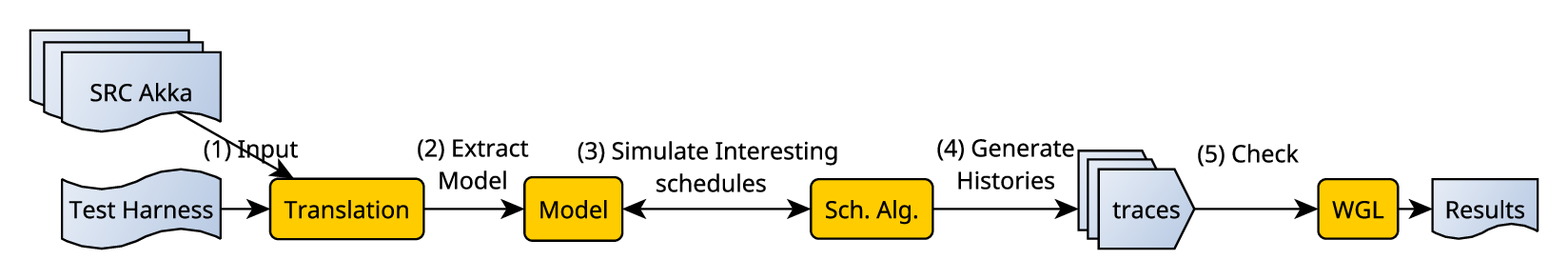}
  \caption{Toolchain workflow. \textbf{WGL} stands for Lowe’s extension of Wing and Gong’s algorithm~\cite{lowe:wgl:linearizability}}
  \label{fig:tool}
\end{figure}


Figure~\ref{fig:tool} shows our setup for checking actor systems.
The user specifies an actor-based system that represents some ADT (e.g.\ a
map) as an instance of our \emph{executable} model.
%

In our approach, each scheduler starts with a set of messages destined
to a set of actors.
For each of these messages we say a destination's \emph{receive} is
\emph{enabled}.
At each step, a scheduler's job is to choose an enabled receive from
the \emph{enabled set}, to execute the action associated with it on the
destination actor, and mark the receive \emph{explored} so that it will not
be revisited from that specific state.
Later, the scheduler may need to \emph{backtrack} to that state so it
explores other interleavings of remaining receives.
Before executing a receive and after marking it as explored, a
scheduler \emph{snapshots} the entire system state and the different
sets of receives.

To start checking an implementation, a user provides the actor system that implements
the ADT, and a set of invocations on it.
Our Schedulers use these harnesses to inject these invocations as messages before the scheduler
starts, then it starts exploring till no more enabled receives remain. 
%
%
Response messages accumulate in a client's queue to be appended to the
schedule, by the scheduler.
The sequence of invocations and responses it discovers form
\emph{histories}\footnote{Note that even schedules that lead to these
histories are accessible in a construct in the schedulers until the
exploration statistics are printed.}.
%

\subsection{Schedulers}

Here, we describe the set of schedulers we compared starting with the
simplest ones. The criteria for choosing these 7 algorithms is
multifaceted: 1) they are popular, 2) all extend each other towards
specialization in a linear inheritance (i.e. to implement LiViola -- LV --  we had to
implement all that it refines/inherits) which reduces the amount
of duplicated code and increases modularity; and 3) they form a basis
for many other algorithms to extend and specialize them as needed.

\subsubsection{Systematic Random (SR)}
From the initial enabled set, the systematic random scheduler chooses
a random receive and executes it.
This may enable new receives to add to the previously enabled ones to
pick from, and the process repeats until no enabled receives remain.
From there, the scheduler backtracks to the initial state and initial
enabled set and repeats until a timeout.
%
\subsubsection{Exhaustive DFS (Depth First Search)}
One straightforward approach to exploring systems is to explore schedules depth
first.
This algorithm non-deterministically picks an enabled receive that has
not been marked explored and performs the associated action.
%
When the enabled set is empty on some path, it outputs a schedule
which later is transformed into a history.
Then, it backtracks to the earliest point in time  where
other, unexplored, receives remain and it repeats this procedure.
To bound execution time, all schedulers have to be stopped at some
point, e.g. at some count of schedules generated.
%
Hence, in practice, this policy will tend to mostly make some initial choices
of receives, and it will aggressively explore reorderings of the ``deepest''
enabled receives before timing out.
As a result, in our experience, this approach tends to explore similar
schedules (before timing out), so it produces many but similar histories.

All of the remaining schedulers are based on Exhaustive DFS.
They cut the search space by overriding methods that refine the
scheduler behavior to prune receives causing redundant
schedules/histories.

\subsubsection{Delay-bounded (DB)}

The delay-bounded scheduler~\cite{delay-bounded} extends the
Exhaustive DFS scheduler, and it mainly explores schedules similarly
but randomly delaying some receives.
The scheduler starts with a fixed delay budget $D$.
As it explores, for each receive $r$, it picks a random natural number d, $0\le d \le D$.
If $d > 0$ then the scheduler skips over $d$ agents that have enabled
receives in the enabled set in a \emph{round robin} order, and
it explores the next agent's enabled receive.
%
The sum of chosen values of $d$ along in a schedule is bounded to
$D$.


\subsubsection{Dynamic Partial Order Reduction(DPOR/DP)}
\begin{figure}[b!]
  \centering
  \includegraphics[width=0.6\columnwidth]{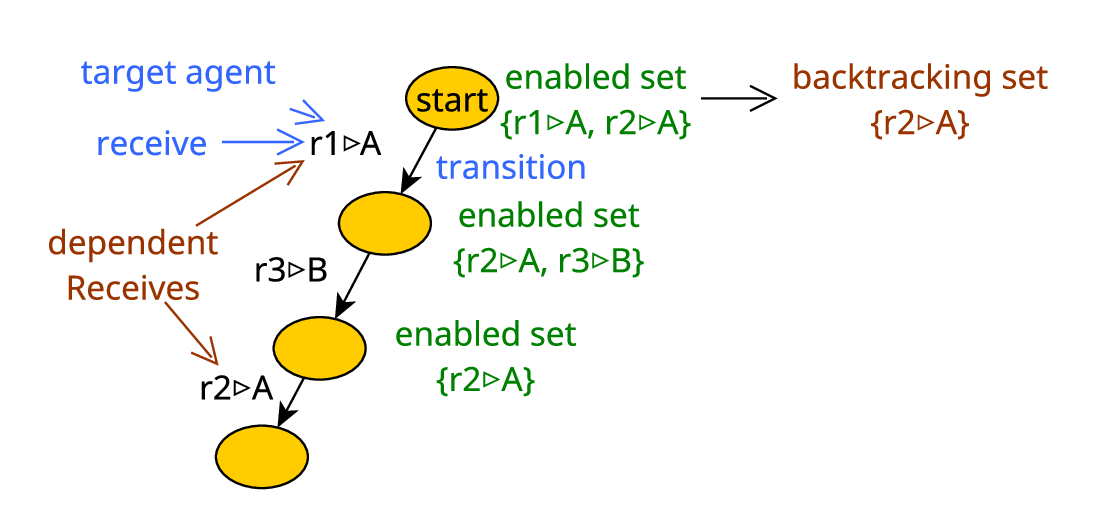}
  \caption{Dependent receives in DPOR.
    \textbf{r1$\vartriangleright$A}: means receive \texttt{r1} heading to actor \texttt{A}.
    \textbf{enabled set}: means the set of receives/messages that can be scheduled/executed next.
    \textbf{dependent receives}: Receives/messages that are sent to (to be executed by) the same actor, per DPOR, TransDPOR definition of dependence. Dependence is redefined later for IRed/LiViola.
    \textbf{backtracking set}: the set of dependent receives upon backtracking to that state, a scheduler can prioritize picking a receive/message from this set to execute/schedule.
  }
  \label{fig:dependent}
\end{figure}


Dynamic partial order reducing (DPOR \cite{flanagan:dpor}) scheduler
prunes schedules that reorder independent receives that affect
different actors.
If two receives target the same actor, then they are considered
\emph{dependent} since the order they are applied in can influence the
behavior of that agent (and transitively those it communicates with).
%
%
In Figure~\ref{fig:dependent}, for example, when exploring a state of
the system, there are two enabled receives; $r_1$ destined to agent
$A$ and $r_2$ destined to agent $B$.
The scheduler picks one of them non-deterministically (here $r_1$),
and it executes the transition, producing $r_3$ destined to $B$.
In the new state, it then chooses to execute $r_3$.
Notice that $r_2$ and $r_3$ are independent so only one interleaving
needs to be explored.
However, when $r_2$ is chosen and executed, the scheduler notices that
the previously chosen $r_1$ is dependent, so it determines that it
must explore the receives in the opposite order as well.
Hence, 
the scheduler first checks if
$r_2$ is enabled in the state 
from which $r_1$ was executed. If so, it adds
it to the first state's \emph{backtracking set}, which tracks the
remaining receives that need be explored after completing the current
path.
It only needs to do costly branching if the receives are dependent.

One complication arises in DPOR (shown in Figure~\ref{fig:dependent})
is that it is possible that $r_2$ is not in the enabled set of the top
state (labeled \texttt{start}).
That is, two receives can be dependent, but they might not always be
in one enabled set.
This situation is due to a later executed receive
\emph{enabling} $r_2$.
In that case, DPOR takes all receives that executed between the
dependent state (i.e. a previous state from which a receive destined
to the same agent as the current receive was executed) and the current
one, and filters them based on whether they were enabled in the
dependent state, returning only those that were enabled in the
dependent state.
Then, the entire set of filtered receives is added to the dependent
state's \emph{backtracking set} to explore later.
%
%
This is where TransDPOR and IRed improve over DPOR, by being
more careful about which one from the filtered set of receives is added
to the dependent state's \emph{backtracking set}.

\subsubsection{TransDPOR (TD)} The key idea in
TransDPOR~\cite{agha:transdpor} that differentiates it from DPOR is
that when a receive becomes enabled that is dependent with a receive
earlier in the schedule, it is added to the backtracking set of that
earlier state (the dependent state) in the schedule \emph{only if}
that state's backtracking set is empty.
TransDPOR always \emph{over approximates} the root enabler receive
(the receive that enabled the current one whether directly or
transitively) by \emph{always} adding the \emph{first receive in the
schedule after the one executed from the dependent state} to the
dependent state's backtracking set.
%
%
\subsubsection{IRed (IR)}

IRed is based on TransDPOR, and improves over it in one specific
aspect.
It tracks causality (i.e. the root enabler receive) with perfect
precision by scanning for the causal chain backwards starting from the
current receive until the first one that enabled it in the schedule.
Then it adds that root enabler receive, if found, to the dependent
state's backtracking set.
Otherwise, if a root enabler was not found, it does not update the
backtracking set.
%
%
More concretely, during the scan, it checks for the predicate (and
keeps track of the \emph{earliest} receive executed in the schedule
that satisfies it) whether the \emph{current receive's sender is the
same as the previous receive's receiver}.
The earliest receive that satisfies it becomes the current root enabler receive.
The scanner continues in this manner until it reaches the dependent state, or the predicate is no more satisfied and the last root enabler receive detected was enabled (co-enabled) in the dependent state.
In which case, the last root enabler receive is the the one to explore when the algorithm backtracks to that specific state.
%
%
The sequence of receives starting at the root enabler up to and
including the current receive is called \emph{causal chain}, and the
root enabler is the first receive in this sequence.
This causal chain pattern is characteristic of actors and purely
sequential communicating processes.
%
%
This \emph{reverse traverse} for the root enabler also filters
away many unrelated receives effectively, and it improves backtracking in
the presence of retries.
%
%
This is where our algorithm thrives in complexity that is common in
distributed systems.
%
%
\subsubsection{LiViola (LV)} Our second algorithm is based on IRed and is
focused on revealing linearizability violations by redefining
the dependence relation of concurrent key-value stores.
Linearizability is
compositional~\cite{pcomp,herlihy:linearizability,linearizability:checker:cpp}.
However, the only time this was exploited for verifying
linearizability is in P-Compositionality work~\cite{pcomp} at the
history level.
In addition, we realized that a linearizability violation is a race
condition on an actor/agent having two interfering receives (due to
network re-ordering) or transitively between the actors/agents
composing the distributed ADT (and, hence, a data race on the
collective state of actors composing the ADT) that is exposed
clients.
We used this fact to restrict the number of interesting schedules that
may produce linearizability bugs by applying it at the schedule
generation stage.
This was done by first augmenting the harness with
additional internal messaging info, and then overriding the dependence
relation to restrict dependent states to those that satisfy all of the
following:
(1) the receive executed from that state has to be targeting the same
agent/receiver (same as before);
%
and (2) the receives have to target a certain key in the distributed
key-value store (compositionality of linearizability).
%
%
So, the hypothesis about our algorithm (LiViola) is that it is
expected to perform the worst in a single key-value store (e.g. a map
that has one key, a register, or a single element set) and perform the
best as more keys are added to the distributed ADT.
Hence, our two algorithms above should thrive on complexity more than
the other algorithms.
In addition, it is noted in WGL paper~\cite{lowe:wgl:linearizability}
that WGL suffers the most when there are more keys; however, our
algorithms should reduce the number of schedules the most when there
are more keys.
In other words, the more WGL has to deal with more keys interleaving,
the fewer schedules our algorithms produce in comparison to others and
the more complimentary they are to WGL checking.

While IRed algorithm's improvement is generic to all problems, LiViola
is specialized to linearizability.
As a result, IRed enables a whole class of algorithms that extend it
and can be specialized in-lieu LiViola to more precisely address
different problems.
That can be done by overriding one/both of the
$\mathtxt{areDependent(receive1,receive2)} \rightarrow \mathtxt{Boolean}$ and
$\mathtxt{didEnable(receive1,receive2)} \rightarrow \mathtxt{Boolean}$ methods.
%


\section{Algorithms in Detail}

In this section, we begin with a visual walk through of the
TransDPOR and IRed algorithms in order to visually spot the
differences.
%
After that, in next section \S\ref{sec:pseudo-code-walk}, we
present the differences between them in the form of pseudocode
walk through to remove any confusion, and to make it easier to code the
algorithms.

\begin{figure}[t!]
  \centering
  \includegraphics[width=\linewidth]{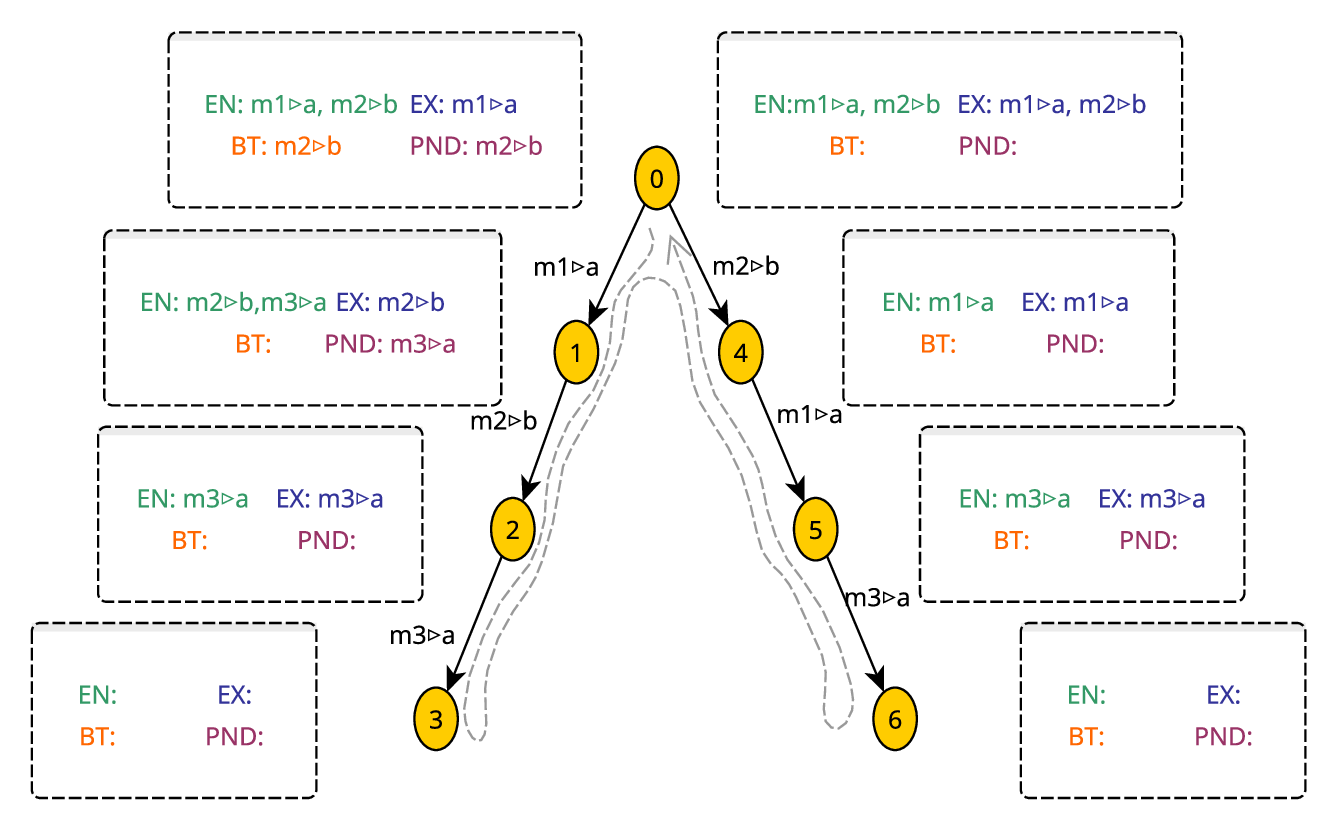}
  \caption{TransDPOR sample run. The dotted line is to show the exploration order in the tree.}
  \label{fig:transdpor:alg:illustration}
\end{figure}

%
\begin{figure}[t!]
  \centering
  \includegraphics[width=.75\linewidth]{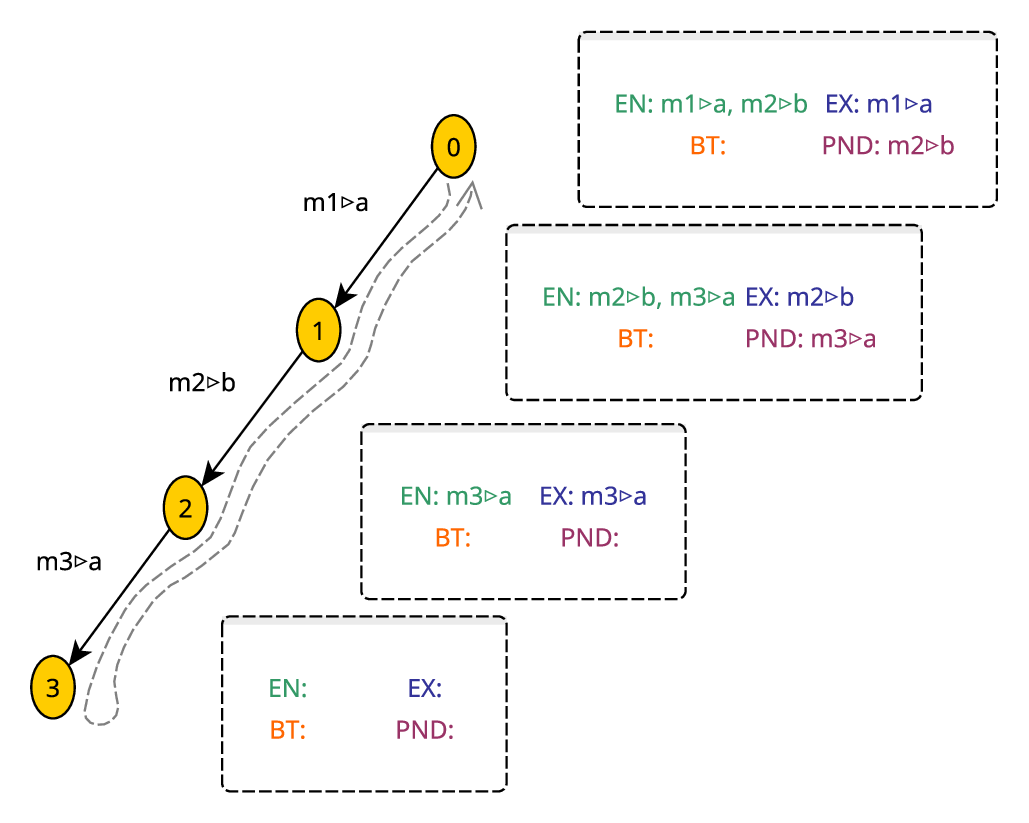}
  \caption{IRed sample run for the same input of TransDPOR in
    Figure~\ref{fig:transdpor:alg:illustration}. The dotted line shows the exploration order.}
  \label{fig:ired:alg:illustration}
\end{figure}

%

Here, we start by explaining the symbols shown in
Figures~\ref{fig:transdpor:alg:illustration} and \ref{fig:ired:alg:illustration}.
Ovals represent states, and arrows represent the transitions
(receives) that are executed from one state and lead to the transition
to the next state.
Different sets are represented with two/three letters: The Enabled Set is
referred to as \mathtxt{EN}, The Explored Set (the \emph{done} set) is symbolized as
\mathtxt{EX}, and the \emph{pending} set is referred to as \mathtxt{PND} (it is $\{\mathtxt{EN}\}
\setminus \{\mathtxt{EX}\}$), and finally the Backtracking Set is referred to as \mathtxt{BT}.
The pending set represents the enabled receives that have not been
explored from the specific state; hence, they are enabled but not in
explored/done set.
It is made explicit to simplify understanding and to relate to the
pseudocode in the next subsection when we explain algorithms in more
detail.
First, we explain TransDPOR visually.
Initially, there are two requests (receives) from two different
clients (shown in the enabled set to the left and right of state 0
in Figure~\ref{fig:transdpor:alg:illustration}), shown in the
enabled set and the pending set is equal to the enabled set.
Once the algorithm randomly picks receive \texttt{m1$\vartriangleright$a}
(i.e. message \texttt{m1} heading to agent \texttt{a}), it
adds it to the explored set (\mathtxt{EX}) and that leads to removing it
from the pending set.
It executes the receive, transitioning from state 0 to state
1, that has the corresponding set shown to its left and enabling
receive \texttt{m3$\vartriangleright$a}, i.e. by executing the action to process
\texttt{m1$\vartriangleright$a} and that action has a send statement that sends out
\texttt{m3$\vartriangleright$a}.
The algorithm proceeds by doing the same, now randomly
choosing to execute receive \texttt{m1$\vartriangleright$b}, and ending at state 2.
It continues the same way but now the algorithm detects that
the \emph{current receive} (\texttt{m3$\vartriangleright$a}) is dependent with a previous
receive (namely \texttt{m1$\vartriangleright$a}); since they are \emph{heading to the same
  agent} (agent \texttt{a}).

Now the algorithm does two checks.
In the first check, it checks if the current receive \texttt{m3$\vartriangleright$a} was
\emph{co-enabled} (resides in the same enabled set of the state from
which the dependent receive was executed, i.e. state 0).
It does not; hence, it was not co-enabled with the receive
\texttt{m1$\vartriangleright$a}, so the algorithm cannot update the backtracking
set of state 0 to contain the current receive.
In the second check, the algorithm goes to try another option to update the
backtracking set (to explore another possible interleaving).
TransDPOR picks the \emph{first receive after the dependent
  receive} (symbolized as $\mathtxt{pre(S,i)}$ in the next section pseudocode)
and \emph{assumes} it is the receive that enabled the current receive
\texttt{m3$\vartriangleright$a} whose sender is agent/actor \texttt{a}.
That receive (\texttt{m2$\vartriangleright$b}) is then added to the backtracking
set of state 0 to be explored upon backtracking to that state.
TransDPOR, at this stage, locks the backtracking set of state 0 
by setting a \emph{freeze} flag, meaning that no more receives are
allowed to be added to that specific backtracking set (that is
part of state 0).
Hence, it keeps the backtracking set size to a maximum of one
receive at any time.
This is how it narrows down the exploration tree in comparison to the
original DPOR that adds more than one receive at a time to a certain
backtracking set.
TransDPOR then executes the transition/receive (\texttt{m3$\vartriangleright$a})
ending in state 3.
At this stage, there are no more pending/enabled receives that
the algorithm can execute, so it backtracks until it reaches a
state where the backtracking set is not empty.
That is, state 0 at this point.

Once it is at state 0, it detects that the backtracking set is
not empty, so it \emph{unfreezes} (sets the \emph{freeze} flag
to \texttt{false}) the state, picks the \texttt{m2$\vartriangleright$b} receive and
executes it transitioning to state 4.
It continues in a similar manner until it reaches state 6, and
after that, it backtracks and exits since no more backtracking sets
are updated.
In the next paragraphs, we explain IRed operation on the same input.

After explaining TransDPOR, we explain IRed algorithm in a similar
manner.
IRed execution of the system proceeds exactly as TransDPOR up until
state 2.
At state 2, it does the first step exactly like TransDPOR does,
i.e. it checks if \texttt{m3$\vartriangleright$a} is \emph{co-enabled} with \texttt{m1$\vartriangleright$a}
in state 0, but it finds it is not.
Then, the second option, it tries to find the \emph{root enabler}
(i.e. the receive that originally, directly or transitively, enabled
receive \texttt{m3$\vartriangleright$a}).
The way IRed does it is different from TransDPOR.
TransDPOR, as we saw earlier, \emph{assumes} that the first receive
after the dependent receive (dependent receive being \texttt{m1$\vartriangleright$a} and
\texttt{m2$\vartriangleright$b} is the one after it in this example) is the one that
enabled \texttt{m3$\vartriangleright$a}.
We know this is not true, and it is an imprecision of
TransDPOR.
IRed, however, is precise at picking the root enabler.
It scans backwards, checking if the sender of receive \texttt{m3$\vartriangleright$a} (we
write it as $\mathtxt{sender}(m3\vartriangleright a)$) is the same receiver of \texttt{m2$\vartriangleright$b} (we
write it as $\mathtxt{receiver}(m2\vartriangleright b)$).
The key idea here is that if \texttt{m3$\vartriangleright$a} was sent by
agent/actor \texttt{b} and \texttt{m2$\vartriangleright$b} was the last receive
received by agent \texttt{b} (same agent), then receiving
\texttt{m2$\vartriangleright$b} could have triggered agent \texttt{b} to send
\texttt{m3$\vartriangleright$b}.
Hence, the root enabler receive becomes \texttt{m2$\vartriangleright$b}.
At this stage, it is not true that \texttt{m2$\vartriangleright$b} is the root
enabler since \texttt{m3$\vartriangleright$a} was sent by agent \texttt{a} not
agent \texttt{b}.
However, we reached the dependent state so the algorithm can not check
the same predicate for the current receive versus the dependent
receive (i.e. $\mathtxt{sender}(m3\vartriangleright a) == \mathtxt{receiver}(m1\vartriangleright a)$) since they are not
co-enabled in the first place and the algorithm has no benefit at
running dependent receive itself, \texttt{m1$\vartriangleright$a}, again from the same
state so it stops before doing so.
If they were co-enabled, however, \texttt{m3$\vartriangleright$a} would have been added
to state 0's backtracking set before the algorithm takes this second
scenario/step of the algorithm.

Elaborating more, assume there was another earlier executed receive
(call it \texttt{m2$\vartriangleright$a}) after the dependent receive, and now the
earlier receive and current one satisfy the predicate above.
The algorithm then makes the earlier receive (\texttt{m2$\vartriangleright$a})
the current one, and it proceeds until it reaches the dependent
receive.
The latest \emph{current receive} tracked by IRed, is then
checked if it was \emph{co-enabled} with the dependent receive
\texttt{m1$\vartriangleright$a}.
If it is, it is added to state 0's backtracking set; otherwise, it
is not added.
To elaborate on the same example, assume there are hundred retries of
the same receive \texttt{m2$\vartriangleright$a}, i.e. the message \texttt{m2$\vartriangleright$a} was sent
a hundred times.
There are two scenarios on how the algorithm deals with this
situation.
The first scenario is that the sender of these \texttt{m2$\vartriangleright$a} retries is
agent \texttt{a} itself, in which case the algorithm picks the
\emph{earliest one}, as we saw in the example, as the root enabler.
The second scenario is that retries of \texttt{m2$\vartriangleright$a} were sent by
another agent, say \texttt{b}, in which case the algorithm will pick
the \emph{latest one}, as in closest to current receive, as the new
current receive (candidate root enabler) and filters away all earlier
ones.
The reason why the other 99 retries are skipped is that they
do not satisfy the IRed predicate.
For example, the 99th \texttt{m2$\vartriangleright$a}
does \emph{not} have the same receiver (agent \texttt{a}) as the 100th
\texttt{m2$\vartriangleright$a}'s sender (agent \texttt{b} in the second scenario), and
such is the case with the rest of retries.
It skips 99 retries, and it resumes from that point on until
reaching the dependent receive and proceeds similar to previously
explained.
Hence, it de-duplicates all these retries that represent a common communication pattern in distributed systems.
Other algorithms, e.g. TransDPOR, would suffer the classic
state space explosion problem in this case.

In the next section (\S\ref{sec:pseudo-code-walk}) we elaborate with
more detail these differences, define all the terms used, and
redefine a few relations to address both differences in TransDPOR,
IRed, and LiViola.
\subsection{The Pseudocode Walk Through}
\label{sec:pseudo-code-walk}

In this section we present IRed pseudocode in detail and will
contrast it to TransDPOR and LiViola. 
%
The algorithm shows the difference between TransDPOR, IRed, and
LiViola.
The differences between TransDPOR and IRed are underlined, while
between LiViola and all others (including IRed) are double underlined.
However, before we can understand the algorithm, we need to explain
some primitives and notations.
We keep most notations the same as in TransDPOR
paper~\cite{agha:transdpor} to simplify understanding.
%
%
\subsubsection{Notations, Terms, and Definitions}
\label{sec:notations:terms:defn}
As explained in our operational semantics paper~\cite{ds2-opsem}, the global \emph{state} of
an actor system is a distributed system state in DS2 terms, and here
it is symbolized as $s \in \mathbb{S}$, where $\mathbb{S}$ is the set
of all possible states in a system.
Each state $s = \langle\alpha,m \rangle$ is comprised of a map
$\alpha: \mathbb{A} \rightarrow \mathbb{L}$ where $\mathbb{A}$ is the
set of all possible agents/actors \emph{identifiers} in the system,
and $\mathbb{L}$ are possible \emph{local states}.
In that state, $m \in \mathbb{M}$ is the set of all pending
messages, while $\mathbb{M}$ is the set of all possible messages in
the system.
We, also, use $pending(s)$ to indicate the set of pending messages for
a state $s \in \mathbb{S}$.

%
Each actor processes each received message atomically
since it does not share any state with any other actor/agent.

That processing step is called a \emph{transition} (or processing a
\emph{receive}).
Processing a transition $t \in \tau$, where $\tau$ is the set of all
possible transitions in a system, may lead to one of the three
outcomes (or a combination of them) depending on the agent's/actor's
local state and constrains imposed by its implementation logic. 
These outcomes are the following: changing its local state, sending
out new messages (we call this enabling new receives/transitions),
and/or creating new agent(s)/actor(s).

\begin{definition}
  \label{defn:transition}
  The transition $t_m$ for a message $m$ is a partial function $t_m:
  \mathbb{S} \rightharpoonup \mathbb{S}$.
  For a state $\langle\alpha,\mu \rangle \in \mathbb{S}$, let a
  \emph{receive} be ($m$,$a$), where $m$ is the message to be received
  by actor/agent $a$.
  Also, let $s$ be the local state of the actor $a$ and $c_a$ be the
  constraint on its local state and its messages, $c_a \subset
  \mathbb{L} \times \mathbb{M}$.
  The transition $t_m$ is enabled if $t_m(\langle \alpha,\mu\rangle)$
  is defined (that is $\alpha(a) = s$ and $m \in \mu$) and $\langle
  s,m\rangle \in c_a$. If $t_m$ is enabled, then it can be executed
  and a new state is produced, updating the state of the actor from
  $s$ to $s^\prime$, sending out new messages, and/or creating new
  actor(s) with their initial state $new_s(t_m)$:
  $\langle \alpha,\mu\rangle \rightarrow^{t_m} \langle \alpha[a
  \mapsto s^\prime] \cup new_s(t_m), \mu \setminus\{m\}\cup
  out_s(t_m)\rangle$.
\end{definition}

Note that the above definition is still verbatim as in TransDPOR
paper, only symbols and the way it was stated differs a little.
The message processed causing the transition $t_m$ be executed is denoted
as $\mathtxt{msg}(t_m) = m$, the actor/agent (also called destination of the
receive) that performed the transition is denoted as $\mathtxt{actor}(t_m) = a$,
the message(s) sent out due to executing the transition is denoted
$\mathtxt{out}(t_m)$, and actors created as $\mathtxt{new}(t_m)$.
Further, we add our definitions to simplify the resulting pseudocode
and make it more understandable.
The sender of a message is denoted as $\mathtxt{sender}(m \in \mathbb{M})$ is
the sender actor/agent of that message.

Next, we also keep TransDPOR~\cite{agha:transdpor} definitions that
follow the standard DPOR~\cite{flanagan:dpor} presentation style.
A schedule (a sequence of transitions) is defined in
Definition~\ref{defn:schedule}.

\begin{definition}
  \label{defn:schedule}
  A schedule is a \emph{finite} sequence of transitions. The set of
  all possible schedules in a system is denoted as
  $\tau^*$ and the execution of a finite sequence of transitions $w \in
  \tau^*$ is denoted by $s \xrightarrow{w} s^\prime$ transitioning the
  system from state $s$ to state $s^\prime$.
\end{definition}

To elaborate on Definition~\ref{defn:schedule}, A \emph{transition
  sequence} $S$ is a finite sequence $t_1.t_2 \dots t_n$ of
transitions where there exists
states $s_0,s_1,\dots,s_n$ such that $s_0$ is the initial state and a
series of transformations of that state to intermediate states
reaching the \emph{terminal state} $s_n$, as in $s_0 \xrightarrow{t_1}
s_1 \xrightarrow{t_2} \dots \xrightarrow{t_n}
s_n$.
Two transitions are said to be \emph{independent} when they are
\emph{not} performed by the same actor/agent.

\begin{definition}
  \label{defn:dependence:old}
  Two transitions are said to be dependent when they are performed by the same
  actor/agent. It follows from that, that the order at which these two receives
  are executed may lead to different local states of the same actor performing
  them. Hence, algorithms need to explore both execution orders.
\end{definition}

All independent transitions can be left without permuting their
execution order, in as much as the system allows, with respect to each
other since they can commute without affecting the resulting state of
their different execution orders.
Two transitions $t_i$ and $t_j$ are denoted $\mathtxt{dependent}(t_i,t_j)$ if and only if
they are \emph{dependent}, per Definition~\ref{defn:dependence:old} for
DPOR/TransDPOR algorithms and per
Definition~\ref{defn:liviola:dependence:relation} for IRed/LiViola.
We also override that by using only transitions' ids to say the same,
as in $\mathtxt{dependent}(i,j)$.
From that, two schedules are considered equivalent based on whether
all that changed between them is the relative re-arrangement(s) of
\emph{independent} transitions.
Definition~\ref{defn:equivalent:schedules} defines this relationship between two
equivalent schedules. It is crucial, at this point, to define the happens-before
relationship in Definition~\ref{defn:happens:before} for the definition of
equivalent schedules (Definition~\ref{defn:equivalent:schedules}) to be clear.

\begin{definition}
  \label{defn:happens:before}
  The \emph{happens-before} relation $\rightarrow_S$ for a schedule $S
  = t_1\dots t_n$ is the smallest relation on $\{1,\dots,n\}$ such
  that:
  \begin{enumerate}
  \item $i \rightarrow_S j$ if $i \le j$ and $\mathtxt{dependent}(t_i,t_j)$
  \item the relation $\rightarrow_S$  is transitively closed.
  \end{enumerate}
\end{definition}

The happens-before relation is the first of two constraints that enforces the
strict ordering of a subset of the transitions in a schedule based on the system
imposed constraints.
This is necessary since it is the basis of partial order reduction;
it imposes partial ordering on some of the transitions during
permuting of schedules.
The second constraint is the direct/indirect enablement between two
receives/messages, which is defined in
Definition~\ref{defn:enablement:transdpor}.

\begin{definition}
  \label{defn:equivalent:schedules}
  Two transition sequences (schedules) $S1$ and $S2$ are equivalent
  if and only if they satisfy both of the following:
  \begin{enumerate}
  \item Contain the same set of transitions.
  \item They are linearizations of the same happens-before
    relations\footnote{We explain what a happens-before relation is next.}.
  \end{enumerate}
\end{definition}

We define some auxiliary functions to be used throughout the rest of the paper:
\begin{itemize}
  \item \textbf{dom(S)} is  the set of identifiers assigned to the events in the schedule/trace \texttt{S} to determine their location in the sequence of events/receives..
  \item \textbf{out(S)} is the messages sent out after executing/processing schedule/trace \texttt{S} events, also overridden for a single event e.g. $\mathtxt{out}(S_i)$.
\end{itemize}

\begin{definition}
  \label{defn:enablement:transdpor}
  In a transition sequence S, the \emph{enablement relation} $i
  \rightarrow_S m$ holds for $i \in \mathtxt{dom}(S)$ and message $m$ if and
  only if one of the following holds:
  \begin{enumerate}
  \item $m \in \mathtxt{out}(S_i)$
  \item $\exists j \in \mathtxt{dom}(S)$ such that $i \rightarrow_S j$ and $m
    \in \mathtxt{out}(S_j)$.
  \end{enumerate}
\end{definition}
The above is TransDPOR's \emph{enablement} relation.
It differs significantly from our (IRed's) enablement definition we
present shortly.
The TransDPOR definition of enablement, shown above in
Definition~\ref{defn:enablement:transdpor}, states that:
1) if a message $m$ is sent from the transition/receive $S_i$, then
that receive \emph{enabled} that message (i.e. enabled the transition
that will process it later when received by the agent it is destined
to), or 2) if an earlier transition/receive $S_i$ enabled another
transition/receive $S_j$ that, in turn, sends a message $m$, then
$S_i$ indirectly (but only through \emph{one intermediate transition}
-- ``exists'') lead to enabling that message $m$.
We need to pause at the second part of the definition.
It allows only \emph{one intermediary} transition/receive to enable and
indirect enablement of a message/receive.
Here is a missed opportunity, i.e. the definition misses partial order
reduction opportunities for coarser grained interleaving.
In other words, TransDPOR definition of enablement relation, does not
capture the full \emph{transitivity} of the enablement relation
defined by IRed.
It approximates to the \emph{first transition after the dependent
  transition} as the root-enabler.
This over-approximation leads to unnecessary additional schedules that
are uninteresting.
This is crucial to take note of as it is the basis of \emph{precise
  causality} tracking of both IRed and LiViola that we define next.

Next we present our new definition of the enablement relation
(IRed's), that is \emph{transitive enablement} relation (``for all'') in
Definition~\ref{defn:enablement}.
It, also, is what we call \emph{causal chains}.
That is, we redefine the enablement relation in the pseudocode to be
that of ours, with minimal change to the original TransDPOR algorithm.
%

\begin{definition}
  \label{defn:enablement}
  In a schedule $S$ the \emph{transitive enablement relation} denoted as $i
  \Rightarrow_S m$ holds for $i \in \mathtxt{dom}(S)$ and a message $m \in
  \mathbb{M}$ if and only if one of the following holds:
  \begin{enumerate}
  \item $m \in \mathtxt{out}(S_i)$
  \item $w \subseteq S$ and $\forall j,k \in \mathtxt{dom}(w)\ |\ i < j < k$ and
    $i \rightarrow_S j$ and $j
    \rightarrow_S k$ and $l = \mathtxt{max}\{dom(w)\}$ and $m \in \mathtxt{out}(w_l)$
  \end{enumerate}
\end{definition}

Note that Definition~\ref{defn:enablement} is \emph{significantly}
different than the way TransDPOR defines the enablement relation.
Specifically, it differs in the second case of the definition.
Here, we define the \emph{transitive enablement} relation based on the
observation that a sequence of transitions each enabling the next in a
certain execution path (sub-schedule or sub-sequence) through the same
or different actors, then that sub-sequence may enable a message/receive
to \emph{eventually but strictly causally} be sent out.
The specific criteria to detect that pattern is specified by a
$\forall j,k$ quantifier over transitions whose indices are $i < j < k$ in
that sub-sequence $w$.
%
That sub-sequence $w$, in turn, conforms \emph{completely} (all
its transitions) to the happens-before relation, it is a total order
by itself.
To recap, the happens-before relation in actors is a relation over
transitions each of which is enabled by a received message (a receive).
This is why we use transitions and receives interchangeably.
This is stated by $i \rightarrow_S j$ for the first transition being
fixed by the relation and for all $j, k \in \mathtxt{dom}(w)$ such that each
$j$ happens-before all $k$'s after it, $j \rightarrow_S k$.
Further, the message $m$ has to be sent out by the last transition $w_l$
($m \in \mathtxt{out}(w_l)$) and whose index is the last in the sub-sequence $w$
($ l = \mathtxt{max}\{\mathtxt{dom}(w)\}$).

%
For better readability of the algorithm presented next, the term $i
\xRightarrow{} m$ can be read as ``the causal-chain that starts with
$i$ and ultimately causes message $m$ to be sent''.
For brevity, IRed's definition of transitive enablement relation means
one transition may enable a series of subsequent acyclic transitions/enablements to
eventually enable (send out) a certain message $m$.
The specific implementation details to detect that pattern will be
discussed in context when explaining the IRed algorithm in the next
section, \S\ref{sec:ired:liviola:in:TDPOR}.

Next, we explain IRed and Liviola algorithms with respect to
TransDPOR in a much similar style as it was done for TransDPOR with
respect to DPOR.
We chose to stick to the same style as it makes understanding the
subtle differences between these algorithms easier to follow, and it
binds previous and current publications all together for better
documentation of advancements.
\subsubsection{IRed and LiViola in terms of TransDPOR}
\label{sec:ired:liviola:in:TDPOR}
In this section we explain the IRed pseudocode shown in
Algorithm~\ref{pseudo:listing:transdpor:and:ired}.
Again, the underlined parts are the differences between the original
TransDPOR algorithm, while the double underlined is the difference
between LiViola and IRed.
We begin by explaining some notations used in the pseudocode to make
explaining the algorithm more streamlined.
\begin{algorithm}[b!]
  \caption{TransDPOR vs. IRed vs. LiViola}
  \begin{algorithmic}
  \item[0:] Initially: $Explore(\phi)$
  \item[]
  \item[1:]
      $Explore(S)$ \{
      \item[2:] \hspace{2em} let $s = last(S)$;
      \item[3:] \hspace{2em} for all messages $m \in pending(s)$ \{
      \item[\underline{\underline{4:}}] \hspace{4em} if $\exists i = max(\{i \in dom(S)\ |\ S_i$ \underline{\underline{is dependent}} and
      \item[]   \hspace{6em} may be co-enabled with $next(s,m)$ and $i \not \rightarrow_S m\})$ \{
      \item[5:] \hspace{6em} if($\neg freeze(pre(S,i))$) \{
      \item[\underline{6:}] \hspace{8em} let $E = \{ m^\prime \in enabled(pre(S,i))\ |\ m^\prime = m$ or 
      \item[]    \hspace{10em} \ $\exists j \in dom(S)\ |\ m^\prime = msg(S_j)$ and
      \item[]    \hspace{10em} $j = min(\{k \in dom(S)\ |\ k > i$ and $\underline{k \xRightarrow{}_S m} \})$
      \item[7:] \hspace{8em} if ($E \setminus backtrack(pre(S,i) \neq \phi)$) \{
      \item[]   \hspace{10em} add any $m^\prime \in E$ to $backtrack(pre(S,i))$;
      \item[]   \hspace{10em} $freeze(pre(S,i)) := true$;
      \item[]   \hspace{8em} \}
      \item[8:] \hspace{6em} \}
      \item[9:] \hspace{4em} \}
      \item[10:] \hspace{2em} \}
      \item[11:] \hspace{2em}if ($\exists m \in enabled(s)$)\{
      \item[12:] \hspace{4em} $backtrack(s) := \{m\}$;
      \item[13:] \hspace{4em} let $done = \phi$
      \item[14:] \hspace{4em} while ($\exists m \in (backtrack(s) \setminus done)$) \{
      \item[15:] \hspace{6em} add $m$ to $done$;
      \item[16:] \hspace{6em} $freeze(s) := false$;
      \item[17:] \hspace{6em} $Explore(S.next(s,m))$;
      \item[18:] \hspace{4em} \}
      \item[19:] \hspace{2em}\}
      \item[20:] \}
  \end{algorithmic}
  \label{pseudo:listing:transdpor:and:ired}
\end{algorithm}


%

%
For a schedule $S = t_1.t_2 \dots t_n$, $\mathtxt{dom}(S)$ is the set of
transitions identifiers $\{1,2,\dots,n\}$, while $S_i$ for $i \in
\mathtxt{dom}(S)$ is the specific transition $t_i$ in that schedule $S$.
The state $s_{i-1} \in \mathbb{S}$ from which a transition $t_i$ is
executed is denoted by $\mathtxt{pre}(S,i)$.
The state $s_n \in \mathbb{S}$ after a schedule $S \in \tau^*$ is
executed is indicated by $\mathtxt{last}(S)$.
%
%
%
%
Finally, we use $\mathtxt{next}(s_i \in \mathbb{S},m)$ to indicate the transition
$t$ that processes the message $m$ starting from state $s_i$.

As we already know, and like DPOR-based algorithms before it, IRed
maintains a \emph{backtracking set} that keeps track of all
receives/transitions that are to be explored from that specific state
$s \in \mathbb{S}$ to which they were added.
However, just like TransDPOR, that backtracking set can only have one
transition/receive at max at all times.
That is implemented using the $freeze$ flag, if it is set in that
specific state, the algorithm does not add anything to that state's
backtracking set, $\mathtxt{backtrack}(s)$.
Otherwise, it adds one transition/receive, and it sets the $freeze$ flag
to prevent more receives from being added, as long as it is frozen.
When the algorithm backtracks to that specific state $s \in
\mathbb{S}$ and $\mathtxt{backtrack}(s) \not= \phi$, it resets the $freeze$
flag and executes the transition/receive from that backtracking set.
Just as a reminder, all our algorithms except Systematic Random are
depth first search algorithms (DFS).
%


Next, we explain (line by line) the pseudocode shown in
Algorithm~\ref{pseudo:listing:transdpor:and:ired}.
IRed starts, like TransDPOR, by finding the current state $s$ for the
input sequence/schedule $S$ (line 2).
The algorithm then loops over all $\mathtxt{pending}(s)$ messages in state $s$
(line 3) and explores them depth first.
Lines 4-10 contain the main logic of the algorithm,
while Lines 11-19 contain the recursive step of the algorithm.
At line 4, it starts by searching for the
\emph{latest} ($\mathtxt{max}\{\dots\}$) dependent
state $\mathtxt{pre}(S,i)$ for the currently being explored transition
$\mathtxt{next}(s,m)$ that are \emph{may/not} be enabled and both are not
governed by the enablement relation $i \not \rightarrow_s m$.
If there is such, the algorithm proceeds to line 5, otherwise it tries
with another message $m \in \mathtxt{pending}(s)$.
If there is no more messages in the pending set $\mathtxt{pending}(s)$, it jumps
to line 11.
Assuming there was such dependent state, however, the algorithm will
then check if the dependent state's freeze flag is \emph{not set}
(i.e. reset), $\neg \mathtxt{freeze}(\mathtxt{pre}(S,i))$.
If it is set, however, the algorithm proceeds to line 11, again.
If the $freeze$ flag is reset, this means the backtracking set of the
dependent state does not have any transition/receive in it, and hence
the algorithm will attempt to find a candidate transition/receive set
to add one from it to the backtracking set, Line 6.
At line 6, the algorithm tries to do one or two checks, depending on
some criteria to be discussed soon, in order to construct the
\emph{candidate receives/transitions set} (indicated by $E$) from
which it adds to the backtracking set of the dependent receive
$\mathtxt{backtrack}(\mathtxt{pre}(S,i))$.
%
%
The first check the algorithm does to find candidate
receives/transitions (or similarly messages) to add to $E$.
To do this, it checks if the current message being explored is
\emph{co-enabled} with the message processed by the transition
executed from the dependent receive, $m^\prime \in \mathtxt{enabled}(\mathtxt{pre}(S,i))$.
%
If there is any, that is added to the candidate set $E$.

Before we explain the second check, there is a note we want to make.
In TransDPOR pseudocode, there was a redundant conjecture of two
terms at the end of the line starting with ($\exists j \in \mathtxt{dom}(S)\ |\
m^\prime \dots$).
We removed that to make it more readable, since they cause
confusion and it is covered by the line that starts with ($j =
\mathtxt{min}\{\dots\}$).
That conjecture was ``$j > i$ and $j \rightarrow_S m$''.

In addition, the algorithm does another check so that if there is a
message that was enabled directly or transitively (indirectly) through
what we called earlier a causal chain, i.e. a series of
receives/transitions for which each earlier transition/receive enables
a later one to be executed.
%
%
The algorithm tries to find a transition $S_k$ that enabled a message
$m$ directly or transitively through later enabled
transitions/receives, $k \xRightarrow{}_S m$.
The said message is found by first extracting a sub-sequence $w$
(Definition~\ref{defn:enablement}), using the transitive enablement relation,
whose all transitions conform to the predicate
$\mathtxt{receiver}(w_{k-1}) = \mathtxt{sender}(w_k)$ and whose last transition
causes the message $m$ to be sent out.
In IRed's implementation, the predicate is used to extract the
sub-sequence $w$ by traversing the transition sequence in reverse (we
call it \emph{reverse traverse}) down to the earliest transition $t_k
\in w$ that satisfies it.
The reason behind this is that the algorithm only knows what
transition is caused by which one that happened before only after
executing all transitions before it.
In other words, the algorithm does not know the future, it only checks the past
transitions to detect the root-enabler $t_j$ where $j \in \mathtxt{dom}(S)$.
That $w$ is a sub-sequence of the schedule $S$, i.e. its transitions
is a subset of $S$ transitions, with the same original relative order in
$S$, that conform to the said predicate.
After the algorithm extracts that sequence, it chooses the first
transition of it as the root enabler of the message $m$, and hence
considers it a candidate to be added to $E$.
All of the above is stated in the pseudocode as: $j = \mathtxt{min}\{k \in
\mathtxt{dom}(w)\ |\ k > i$ and $k \xRightarrow{}_S m\}$.
%
%
%
%
%
%
%
%
%
What was just explained is the more precise tracking of the
\emph{root enabler} (i.e. the first transition in the sub-sequence
$w$) that IRed tracks precisely in comparison to TransDPOR's redundant
over-approximation of the root enabler (the first transition/receive that
happened after the dependent transition/receive in $S$).
%

Now, the only difference between LiViola and IRed is that LiViola
overrides the $\mathtxt{dependent}(i,j)$ relation and tightens it a bit more
than just \emph{two receives/transitions are dependent if they are
  executed/performed by the same receiver actor/agent}.
It is double underlined in the pseudocode shown in
Algorithm~\ref{pseudo:listing:transdpor:and:ired}.
Definition~\ref{defn:liviola:dependence:relation} redefines the
dependence relation for LiViola.

\begin{definition}
  \label{defn:liviola:dependence:relation}
  Two transitions/receives are dependent if and only if they satisfy
  all of the following:
  \begin{enumerate}
  \item They are performed by the same actor/agent
  \item They are affecting the same variables in the local state of
    the actor/agent
  \item They are not constrained by a happens-before or (transitive) enablement relation
  \end{enumerate}
\end{definition}

Lines 11-19, are the recursive step of the algorithm.
In line 11, the algorithm checks if the message under investigation
$m$ (that was picked randomly from the pending set) is in the enabled
set of the last state in the schedule, $\exists m \in \mathtxt{enabled}(s)$.
If not, it tries with other messages $m \in \mathtxt{pending}(s)$ looping back
to Line 3.
%
If the message is in the enabled set of that last state $s$, the
algorithm proceeds to update the backtracking set of $s$ to the
randomly picked message $m$, $\mathtxt{backtrack}(s) := \{m\}$ (Line 12).
The reason behind this is that the algorithm always checks the
backtracking set for the next message to explore, it simplifies the
implementation.
At line 13, the algorithm resets the new state's explored set $done$
so that it marks future to-be-explored receives/transitions.
The loop in line 14, then, recurses over all messages that are in the
backtracking set of the latest state but that were \emph{not} explored
before, $\exists m \in (\mathtxt{backtrack}(s)\setminus \mathtxt{done})$.
Each message $m$ is marked as explored, $\mathtxt{add}\ m\ to\ \mathtxt{done} $ (line
15).
Then the state is marked as unfrozen, i.e. the algorithm can update
its backtracking set by future transitions/receives, $\mathtxt{freeze}(s) :=
\mathtxt{false}$ (line 16).
Finally, the algorithm recurses on the next state resulting from
processing the message $m$ (as in performing the transition/receive),
$\mathtxt{Explore}(\mathtxt{S.next}(s,m))$, and appending the resulting transition to the
transition sequence/schedule $S$ (Line 17).
The algorithm continues until there are no more
messages in the pending set.

The next section will detail our experiments for evaluating the
performance of all the mentioned algorithms.

\section{Evaluation}
\label{sec:evaluation}
Since the primary goal of this work is to assess the effectiveness of
said schedulers, we have devised five different actor systems.
We detail them in the next subsections, use the tool-chain to find
linearizability violations, and compare the various algorithms' performance in finding violations.

\subsection{Correct Distributed Register (DR)}
\label{sec:corr-distr-regist}
The simplest actor system we test with is a register ADT
(\texttt{read()}, \texttt{write(v)}) which is primary-backup
replicated (Algorithm~\ref{pseudo:listing:dist:register}).
One agent is statically designated as primary, and the others are
backups (lines 1-5).
When any replica receives a \texttt{write(v)} message it forwards it
to the primary/leader (lines 15-17).
When the leader receives the \texttt{write(v)} message, it processes it
(lines 11-14) by initiating a count for writes acks (counting
self), broadcasting \texttt{WriteReplica}, waiting
for the \texttt{WriteReplicaAck}'s received to reach a majority
quorum (lines 30-34), and then sending a \texttt{WriteAck} to the
client if majority acks was reached.
When replicas receive \texttt{WriteReplica} messages (lines 25-27),
they write the value to the register and send back a
\texttt{WriteReplicaAck} to the sender (the leader).
Reads are processed similarly but with their respective protocol
messaging.
Algorithm~\ref{pseudo:listing:dist:register} has all the remaining
details.
If some backups disagree about the current register state, and the
acks count reaches a full count without quorum agreement, the primary
retries the operation.
However, we discovered that our implementation of retries is actually
dead code (never executes and hence not shown here).
%
%

\begin{algorithm}[t!]
  \caption[Correct Dist. Reg. Pseudo]{Correct distributed register
    pseudocode per agent in the ADT}
  \begin{multicols}{2}
    \begin{algorithmic}[1]
      \Require{At least 2 agents in the system}
      
      \If{id == 1}
      \State{be leader}
      \Else
      \State{leader $\gets$ 1}
      \EndIf

      \Repeat
      \State x $\gets$ receiveMessage

      \If{x is IAmLeader}
      \State leader $\gets$ x.sender
      
      \ElsIf{x is Write}
      \If{isLeader}
      \State initWritesAcksTracker(value,client)
      \State reg $\gets$ value
      \State broadcast WriteReplica(reg) to peers
      \Else
      \State forward x to leader
      \EndIf

      \ElsIf{x is Read}
      \If{isLeader}
      \State initReadAcksTracker(reg)
      \State broadcast ReadReplica to peers
      \Else
      \State forward x to leader
      \EndIf 

      \ElsIf{x is WriteReplica}
      \State reg $\gets$ x.value
      \State send WriteReplicaAck(x.params) to x.sender

      \ElsIf{x is ReadReplica}
      \State send ReadReplicaAck(x.params, reg)

      \ElsIf{x is WriteReplicaAck}
      \State update writes tracker with x.params
      \If{reached write majority acks}
      \State send WriteAck to x.client
      \EndIf 

      \ElsIf{x is ReadReplicaAck}
      \State update reads tracker with x.params
      \If{reached read majority acks}
      \State send ReadAck(reg) to x.client
      \EndIf 

      \EndIf 
      \Until{Agent Stops/Killed}

    \end{algorithmic}
  \end{multicols} 
  \label{pseudo:listing:dist:register}
\end{algorithm}
\vspace{-1em}



\subsubsection{Clients} Synthetic clients created by schedulers are empty
agents without any behavior.
The scheduler creates a client per request.
Each client submits one request in its lifetime and receives
at most one response.
The scheduler then picks these responses and appends them to the
schedule at hand.
After done generating all schedules, the accumulator construct that
tracks these schedules and their associated data statistics
generates the histories from the schedules.
%
%
It is up to the construct implementation that is
accumulating schedules to decide how to use these schedules as a
post-processing step to transform these schedules.

\subsubsection{Harness Details}

Beyond the actor system itself, linearizability checking requires some
set of client invocations of the ADT methods so that algorithms can
observe the outcomes and produce histories to check.
Different patterns of client requests have different trade offs.
Simple harnesses with few invocations may not produce buggy histories,
while complex harnesses with many invocations may suffer state space
explosion.
Hence, we explore a few small harnesses chosen for each specific ADT
that attempts to mix method invocations that observe state with those
that mutate it.

Linearizability is defined on what clients observe, so harnesses only make
sense if they generate multiple invocations that observe the target
state.
Furthermore, bugs are most likely to manifest when those observations
could have been affected by interfering mutation operation(s).
For the register, we start with a simple harness
that generates two \texttt{read()} operations and a \texttt{write(v)}
operation (we call this harness the \emph{2r+1w} harness).
We also use a harness that includes a second \texttt{write(v$^\prime$)} operation (\emph{2r+2w}), since
concurrent mutations are often a source of bugs.
%
%
The latter case with mutating operations can timeout, so it
finds \emph{fewer} bugs than first harness.

Each harness has six sections.
The first specifies the initial state of the
sequential specification and
whether the ADT is a MAP or a SET (a
register is a one-key key-value map).
The second section specifies sets of target agents' identifiers
in order to distinguish them from other agents
(e.g.\ synthetic clients).
The next three sections indicate how the agents should be initialized,
and it provides patterns that bind messages to message categories that
the scheduler can understand. For example, it provides patterns that
let it recognize read, write, and replication messages and their
acknowledgments in agents' queues.
%
%
The final section is provided specifically for LiViola to indicate which
messages are of interest to interleave, where the key lies in the
payload of these messages (if it is not known until runtime, then a
wildcard of -1 can be provided), whether it is a
read-related/write-related/both message.
The final section also contains some exclusions for messages; only LiViola
respects these; it does not shuffle these excluded messages with rest, which
helps it control state space explosion.
The messages should only be excluded from interleaving if all the following conditions apply:
\emph{receiving that message at the destination}
\begin{itemize}
\item should \emph{not} \emph{interfere} or cause \emph{potentially
interfering messages to be sent} back into the ADT cluster (to ADT
agents);
\item should \emph{not} \emph{mutate agent state} (e.g. change the key-value pair);
\item should \emph{not} change the \emph{observed value at the client}
(even if it blue change the state in question).
\end{itemize}
That being said, we could not apply any exclusions to the correct
distributed register harness.
Also, note that excluding a message does not mean that it is not executed,
LiViola just does not shuffle it; it executes in whatever order it occurs.
These exclusions are both \emph{problem specific}
(i.e. linearizability in this case) and \emph{implementation
specific}.
%
The source code gives the precise format and details of the harness format.

\subsection{Buggy/Erroneous Distributed Register (EDR)}

Our ``buggy'' distributed register is nearly identical to the correct
distributed register, except it does not wait for majority agreement among backups
before it acknowledges a \texttt{write(v)} operation to a client
and some more buggy behaviors, e.g. randomly generated values from
thin air.
This can lead to an acknowledged write operation that is not observed by
a read operation that started after it, violating linearizability.
%
The write operation starts and completes; subsequently a read gets
issued, but it does not observe the value that should have been
installed in the register by the completed write.
We use this buggy register to make sure the various schedulers are
effective at finding linearizability violations in a timely
manner.
The quorum is fixed to only \emph{two} acknowledgments.
Also, we raised the number of agents to \emph{three} for bugs to
manifest.
We use the same harnesses for this case as we do for the correct one
except for one change; we remove the exclusions since it no longer satisfies
the criteria.
\subsection{Another Distributed Register (ADR)}
We implemented another, more complex, distributed register where all
agents can issue read and/or write requests.
%
%
This simply extends the correct register so that all agents act as a leader,
except that concurrent operations some cause retries in the agents to repeat replication operations that overlapped from competing writes.
%
%
%
%
%
This implementation of is similar to Paxos~\cite{paxos} in operation.
%
\subsection{Open Chord (OC)}
\begin{figure}[b!]
  \centering
  \input{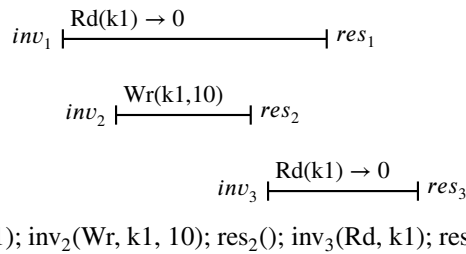}
  \caption{A buggy
    history found by LV in Open Chord Benchmark.}
  \label{fig:buggy:chord:history}
\end{figure}

%
Open Chord~\cite{chord} is a peer-to-peer scalable and performant
distributed hash table (DHT).
It is based on a ring topology of communicating agents that distribute
the load of keys and their values using a consistent hash function.
The simplest form of Open Chord ring is a single agent.
Then another agent/node can join by finding who is its successor based
on a hash of its identifier.
Each node can have backups for other nodes in case it goes missing, we
do not implement this since we do not handle agents
leaving/going missing in this work.
%
%
We implemented the \emph{joining} of nodes (agents) into the ring.
During that process and during normal operation of Chord processes
can join and/or leave the ring.
Since nodes may join during normal operation and they are detectable
by clients prior to completely joining the ring, we made the harness
and startup sequence send the joining message \texttt{FindSucc}
interleave with client requests.
Our tool finds linearizability violations;
one violating history that it
finds is illustrated in Figure~\ref{fig:buggy:chord:history}.

%
Listing~\ref{lst:chord:err:schedule} shows the schedule that explains
and led to that history.
The problem that the schedule shows is that $N_2$ has not finished
joining the ring before it receives a request to read key $k_1$.
Since it believes it is its own successor by default, it replies to the clients without knowing that a newer value for $k_1$ has been stored at $N_1$.
%
%
%
Zave's work~\cite{make-chord-correct} also explores similar correctness issues
that arise in the original Chord specification.
\newline
\linebreak
\noindent\begin{minipage}[center]{1.0\linewidth}
\begin{lstlisting}[caption = {This schedule (i.e. message exchanges
between agents) is what produced the buggy (non-linearizable) history
shown in Figure~\ref{fig:buggy:chord:history}. Initially, \texttt{N1}
is the only node in the ring, then node \texttt{N2} starts joining by
sending a \texttt{FindSucc} message but meanwhile there are few reads
and writes happening. The second read issued by client \texttt{IR1}
(sent to \texttt{N2}) executes before \texttt{N2} is fully joined into
the ring (i.e. before step 9) reading a stale value of key \texttt{k1}
(specifically at step 6), which causes the linearizability
violation. The subscripts track which agent initiated the message, or
its subsequent messages.}, label = lst:chord:err:schedule,
mathescape=true]
1:IR2 $\xrightarrow{Read(k1)}$ N1;2:IR0 $\xrightarrow{Write(k1,10)}$ N1;3:N2 $\xrightarrow{FindSucc_{N2}}$ N1;4:N1 $\xrightarrow{IntWriteResp_{IR0}}$ N1;
5:N1 $\xrightarrow{WriteResp}$ IR0;6:IR1 $\xrightarrow{Read(k1)}$ N2;7:N1$\xrightarrow{IntReadResp_{IR1}}$ N1;8:N1 $\xrightarrow{ReadResp(0)}$ IR2
9:N1 $\xrightarrow{SuccFound_{N2}}$ N2;10:N1 $\xrightarrow{UpdatePred_{N2}}$ N2;11:N2 $\xrightarrow{IntReadResp_{IR1}}$ N2;12:N2 $\xrightarrow{ReadResp}$ IR1
\end{lstlisting}
\end{minipage}

\subsection{Paxos-replicated Map (PX)}

Finally, our most complicated example is a Multi-Paxos-replicated
key-value map.
Each actor of the system maintains an ordered log in the style
of standard state machine replication-based
approaches~\cite{viewstamped.replication.vs.others}.
Client \texttt{write(k, v)} requests are replicated into a log via
majority quorum using Paxos.
A designated leader handles \texttt{read(k)} operations directly
returning the most recent $v$ associated with $k$ among the write
operations recorded in its log after being voted by majority quorum.
The intuition behind Paxos is to keep a monotonically increasing
\emph{proposal id} (i.e. transaction id) to make sure it only
processes and commits changes by the ``latest'' operation
initiated.
Those operations that were initiated earlier but
interfered with later ones, may get rejected in the first phase of the
algorithm, retried with a later \emph{proposal id} till they go
through to the second phase, then they get committed to the logs by the
vote of majority quorum acceptance.
After this, a client gets a response to its request (invocation).
%
%
%
%
For more in depth information about Paxos, please refer to the
literature by Lamport~\cite{paxos,paxos-simple}.

\subsection{Other Systems}
Unfortunately, we ran out of time to port and debug a few more
interesting systems we wanted to explore.
Overall, our tool is suited to similar replication algorithms.
For example, we have a mostly-complete port of Zab~\cite{zab} that we plan to
check with, which is available with our tool's source~\cite{ds2:repo}.
We also experimented with Raft~\cite{raft}, but we were unable to find
reliable Akka implementations for it, so we leave it for future work.
%

\subsection{Metrics and Methodology}

We benchmark each of the above systems with each scheduler from
Section~\ref{sec:overview}, averaged over three runs since some of the schedulers are non-deterministic (e.g.\ Delay-Bounded).
All schedulers share a substantial amount of code and mainly override
behaviors on the Exhaustive DFS, so differences in
runtime are mainly due to real algorithms differences.

\begin{figure}[b!]
  \centering
  \begin{minipage}[center]{0.9\linewidth}
    \centering
    \input{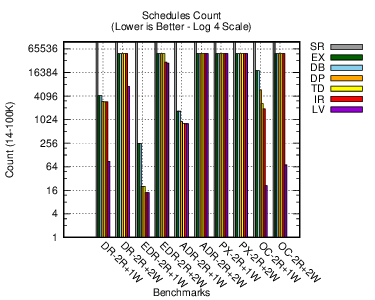}%
    \input{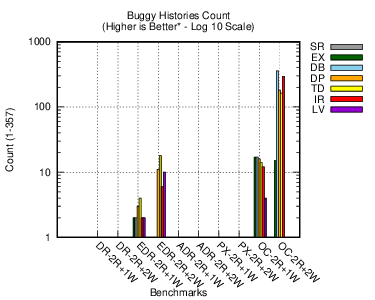}%
  \end{minipage}
  \begin{minipage}[center]{0.9\linewidth}
    \centering
    \input{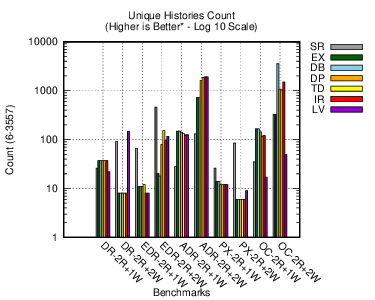}%
    \input{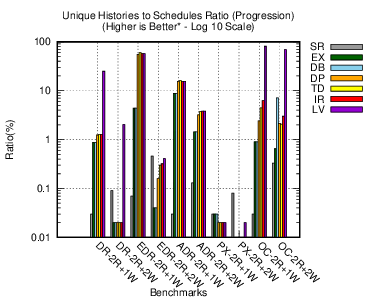}
  \end{minipage}
  \begin{minipage}[center]{0.9\linewidth}
    \centering
    \input{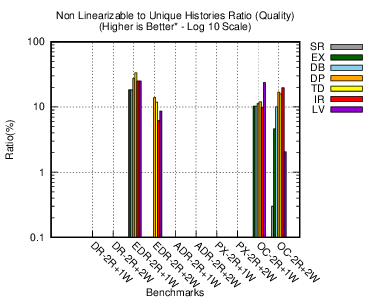}%
    \input{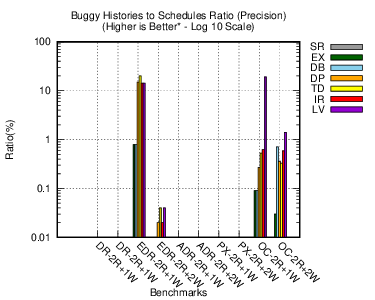}%
  \end{minipage}
  \caption{The First Half of Performance Numbers Summaries for All Algorithms and
    Benchmarks}
  \label{fig:all:histograms:1}
\end{figure}

\begin{figure}[t!]
  \centering  
  \begin{minipage}[center]{0.9\linewidth}
    \centering
    \input{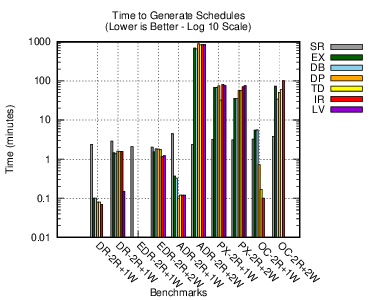}%
    \input{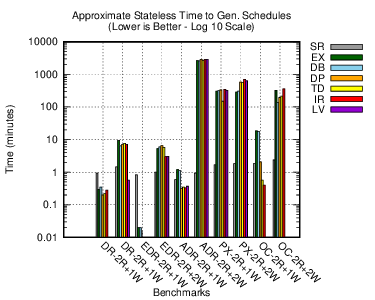}%
  \end{minipage}
    \caption{The Second Half of Performance Numbers Summaries for All Algorithms and
    Benchmarks}
  \label{fig:all:histograms:2}
\end{figure}

Figures~\ref{fig:all:histograms:1} and \ref{fig:all:histograms:2} give
the main results, which we step through in detail in the coming
subsections; we describe the most important metrics here.

%

We call a specific combination of scheduler, actor system, and harness a
\emph{configuration}.
Each schedule leads to a \emph{history} of invocations and responses.
%
We say a history is \emph{unique} if, for a given configuration, no
other history records the same events/entries (invocations and
responses) in the same order with the same arguments, senders,
receivers, and return values.
That is, they are only \emph{chronologically} unique.
So, there could be some histories deemed unique but they are repeated
many times except for re-arrangements of some entries that do not
reveal a violation; those independent receives that materialize to a
history's entries.
%
%
A history is non-linearizable if it cannot be generated by a
linearizable implementation of the ADT being checked (e.g. a linearizable
register or a map).

For a given configuration we call the ratio
of non-linearizable
histories to unique histories that are produced
($\mathtxt{NL}/\mathtxt{UH}$) the \emph{quality} of that configuration.
A high quality means that this configuration produces many examples of
bugs while avoiding the need to check a large number of histories.

Similarly, we call the ratio
of unique histories produced to schedules
explored 
($\mathtxt{UH}/S$) the \emph{progression} of that configuration.
Intuitively, a high progression rate means the scheduler of that
configuration produces a diverse set of histories with little
exploration.

Finally, we call the ratio
of non-linearizable histories produced to
the number of schedules explored
($\mathtxt{NL}/S$) the \emph{precision} of a configuration.
A configuration with high precision finds bugs by exploring fewer
schedules, avoiding wasted work in fruitless ones.

For faster reference, all the above symbols and benchmarks abbreviations and their
descriptions are shown in Table~\ref{tab:legend}, while the
raw numbers for the benchmarks results are shown in Tables~\ref{tab:dr1}
through~\ref{tab:oc2}. Each two consecutive tables starting from
Table~\ref{tab:dr1} show results for the same benchmark but one for 3-receives
and the other for 4-receives.

As we will show later, our results show that the short histories that our harnesses produce mean that history checking times are low.
Ultimately, this means that for our harnesses, good progression is crucial
and good quality of the discovered histories is less important.

We record two different times for schedule exploration.
The default is the time to generate the schedules in a \emph{stateful}
manner; meaning, the scheduler keeps snapshots of the actor system's
state during scheduling.
The second one is an approximation of a \emph{stateless} exploration
of the system.
This is as if the system is restarted after generating each schedule
and controllably reconstructs different schedules in different runs,
visiting different states than previously visited by earlier
schedules.
%
The advantage is memory pressure reduction during exploration.
%
However, the stateful approach is faster to the finish line (overall
time) at the expense of capturing states (using more memory) and
slower per-iteration compute time.
The reason why the stateful schedulers are faster overall time is that
the prefixes of schedules are not executed as often as in the
stateless schedulers.
So, that saves a lot of time for stateful schedulers.

For most systems, we can explore the most interesting cases with just a
few agents, so we only use two agents to test all but the buggy
register, for which we use three agents.
This still has the possibility of producing all of the externally
visible system/ADT behaviors.
%
More complex protocols could require more agents to explore all behaviors;
for example, agent join/leave in protocols like Chord could require
more agents to ensure all internal behaviors are exercised during
schedules generation.
Our implementation, however, does not need more as we do not have
leaving processes and/or dropped messaging; we only implement
joining.

Finally, with enough agents and invocations these algorithms can run
for exceedingly long periods of time.
We terminate exploration after 50,000 schedules per-configuration and
then check the resulting histories.
This cutoff means schedulers may miss fruitful parts of the
exploration, and we observed such situations.
%
%
We tried to strike a balance in choosing this cutoff, but no single
cutoff is likely to work well for all configurations.

%
Last, the specific benchmarks numbers are as follows.
The Lines Of Code (LOCs) in each benchmark range from few hundred
lines to around a thousand lines.
The number of actions involved range from 6 to 9 actions, not counting
the start action that starts the agent.

All bugs found are real bugs, and all of the benchmarks were buggy (both
linearizability and other bugs) before we debugged them.
We kept all previous iterations of them as well in the
repository~\cite{ds2:repo}.
All of the benchmarking was done on the same machine, with each scheduler being single
threaded.
The specs of the machine are the following: Dual 3.5 GHz Intel XEON E5-2690 v3, 192GB
DDR4 ECC, HP Z840 workstation.

A final note, we stressed tested IRed on the heaviest (buggy)
implementation we have, namely the ADR register, for over 203 hrs
monitoring it using VisualVM profiler and took notes of 20-25\% of
non-blocking single threaded CPU usage, 0\% of Garbage Collector (GC)
CPU usage all the time, and 4 GB of average use of heap memory and a
max and min of 5 GB and 3 GB, respectively.

\subsection{Results and Analysis}
\label{sec:results-analysis}  
Figures~\ref{fig:all:histograms:1} and \ref{fig:all:histograms:2} 
show the results of running the benchmarks
across all of the configurations we described.
We work through several dimensions of the table to highlight the key
insights from the results.

\subsubsection{Low History Checking Times}
In virtually all cases, configurations avoid an explosion of runtime in
checking histories using the exponential WGL algorithm ($\mathtxt{TC} \approx
0$).
Some of the configurations produce several unique histories ($\mathtxt{UH} \in [6,
3557]$, Figure~\ref{fig:unique:histories:histogram}), but since the harness
is constrained to just a handful of invocations, checking all of them is
still nearly instantaneous ($\ll$ 1~second).
Hence, though we were worried the complexity of checking histories
would be problematic (poor quality), the explosion in schedulers state
space seems to be a more serious issue (poor progression)
(Figure~\ref{fig:schedules:histogram} and \ref{fig:times:histogram}).

This suggests, just as in conventional model checking, \emph{smarter
pruning or direction of scheduling is more important in finding bugs
than reducing history checking time}.
This also reinforces our decision to focus our efforts on improving
schedulers for linearizability checking rather than focusing on
improving linearizability checking itself as others have
done~\cite{pcomp}.

\subsubsection{Systematic Random}
\label{sec:syst-rand-perf}

The Systematic Random scheduler is ineffective in finding bugs (Figure~\ref{fig:buggy:histories:histogram}).
It progresses well, and it produces many unique histories quickly (Figure~\ref{fig:unique:to:schedules:histogram}).
However, even after doubling its cutoff to let it explore more
schedules, it still finds only one bug in one configuration (the second
harness of Open Chord), where other approaches find many more.

Coupled with the previous conclusion, this suggests that simply
generating more histories alone is not sufficient to find bugs.
This also suggests that any approach that simply tries to maximize
``coverage'' in the space of schedules or in the space of histories
is not likely to yield bugs unless it is efficient enough to provide
near-full coverage, which is unlikely due to the exponential explosion
of exploration space.

\subsubsection{Exhaustive and Delay-Bounded Performance}
\label{sec:exha-delay-bound}

Exhaustive DFS (EX) and Delay-Bounded (DB) perform similarly in most cases
and metrics, even though DB uses some bounded randomness.
By delaying events, DB reorders the space of schedules, but without a cutoff, EX
eventually explores the same schedules in a different order.
So, DB's limited randomness does little to change the set of schedules
explored.
 
Both approaches do well at finding bugs with the first harness (2r+1w).
In the other harness (2r+2w), they hit the cutoff rather quickly (Figure~\ref{fig:schedules:histogram}),
which indicates they generated too many fruitless schedules from the
state space.
This is reflected in their poor precision ($\mathtxt{NL}/S$, Figure~\ref{fig:buggy:to:schedules:histogram}) and progression
($\mathtxt{UH}/S$, Figure~\ref{fig:unique:to:schedules:histogram}) ratios.
If we removed the cutoff, they will find bugs in the second benchmark
but only after generating large numbers of uninteresting schedules,
which indicates they are not effective exploring larger state spaces.
 
%
%
%
%
%

An important observation is that \emph{chronological uniqueness} of
histories (involved in many measures), indicated by an asterisk `*' in
graphs of Figures~\ref{fig:all:histograms:1} and
\ref{fig:all:histograms:2}) causes an issue.
It causes redundancy in the counts of unique histories
particularly so for schedulers exploring more redundant schedules
(e.g. SR and DB).
It amplifies the \emph{illusion} of their
effectiveness in generating more unique histories.
The following is the list of all schedulers, ordered from most to
least affected by this issue: SR, DB, EX, DP, TD, IR, LV.
As we go from left to right in that list, the less amplification
effect we get because there are fewer repeated chronologically unique
histories; hence, the more credible the measures are of that
specific scheduler.
LiViola is the best due to being the least redundant.
%
%
%
Unique histories it produced tended to have the most diversity among
all.
%
%
Even with this diversity, in Open Chord benchmarks, for example, LV
produced four non-linearizable histories out of 17 unique histories on
the smaller harness.
When we cross-checked them with the larger harness we saw that they
are the same history (i.e. the same exact bug), shown in
Figure~\ref{fig:buggy:chord:history}, and explained by the schedule
shown in Listing~\ref{lst:chord:err:schedule}.
Note that the other write call of the second harness is on a different
key ($k_2$); hence, not shown, since it does not affect LV.
However, it does affect all others' results.
If the $k_2$ related invocation and response are placed at the end of
the history, one can imagine how many internal schedules that can be
repeating the bug before it.

%
%
Another example on the other (most affected) extreme in Open Chord
benchmarks, DB produces 3,557 unique histories and that reduce to
357 buggy ones while EX (which covers the same exhaustive state space),
produces 325 unique histories that reduce to 15 buggy histories.
None of them is more exhaustive than the other but DB is more
repetitive than EX.
The reason behind this is when the bug happens to be closer to the
root of the exploration tree and the scheduler is interleaving things
closer to leaves, the prefix containing the bug repeats as many times
as the leaves.
That has an amplification effect on bugs reported especially when the
later interleavings after the prefix result in more chronologically
unique histories.

\subsubsection{DPOR, TransDPOR, and IRed Performance}
\label{sec:dpor-perf}
In the distributed register, the three algorithms DPOR, TransDPOR, and
IRed did better than EX and DB on both harnesses.
DPOR, however, did slightly less favorably than TD and IR in terms of
timing. 
The latter two did exactly the same.
Otherwise, IR did not show any improvements over TD because this
implementation is strongly causally consistent; hence, IR does not
get an advantage over TD's over-approximation of the root enabler.
So, TD represents the worst-case scenario for IR.
%

%
In the second benchmark, DP and TD did mostly the same except for few
differences on the 3 invocation test.
DP produced fewer \emph{quality} unique histories to catch bugs as
indicated by $\mathtxt{NL}/\mathtxt{UH}$, and it is less \emph{precise} as indicated by
$\mathtxt{NL}/S$ than TD.
However, results show it is more precise than IR.
This is tempered by what we mentioned regarding repetitive
chronological uniqueness of histories.
TD was significantly faster at making \emph{progress} towards unique
schedules than DP.
As for IR in the second benchmark, precision, quality and
progression are slightly less than TD, but TD is more
repetitive.
%
%

\subsubsection{LiViola vs IRed Performance}
\label{sec:liviola-vs-ired-perf}
LV performance numbers should have been exactly like IR's on the
register benchmarks, since they are one-key stores.
%
%
LV
enables developers to tweak the harness with interleaving exclusions (within the constraints we
mentioned in \S\ref{sec:corr-distr-regist} to assure soundness) to
indicate which messages are the focus on during exploration
(i.e. potentially interfere) which lead to significant improvements.
It explores only 88 schedules in comparison to the second
highest runners' (IR and TD) 2,906 on the smaller harness, and
does not approach the cutoff on the 4-invocations test, at 7,236
schedules.
Meanwhile, IRed ran for over 203 hrs and still never terminated.
%
%
However, developers should practice caution using this feature as we
will see why in the next benchmark.

In the second benchmark, IR and LV were the best of the algorithms,
too.
They performed similarly except LV scored significantly better but were
2 seconds slower.
In the third benchmark, LV performance was the worst-case scenario
showing exact statistics as IR.
Note that on the buggy version of this third (not shown here), when
we tweaked the harness aggressively, LV outperformed all others by a large
margin in the first harness.
However, for the second test, it missed all bugs and prematurely
terminated at a bit over 10K schedules stating it did not find bugs;
when it should not miss bugs.
%
%
So, we reverted these tweaks and passed a plain harness, i.e. with no
tweaks, and re-run the experiments, shown in the
Figures~\ref{fig:all:histograms:1} and \ref{fig:all:histograms:2}.
This is an example of why developers are to practice caution when
tweaking the harness while still conforming to the criteria presented
in \S\ref{sec:corr-distr-regist}.
%
The reader is encouraged to look at the rest of the numbers in tables
in Appendix~\ref{benchmarks:appendix} for raw numbers, the best
results we observed are for Open Chord.

\subsection{Limitations}
\label{sec:limitations}
The model still has a limitation in conditionals e.g. \texttt{If} and \texttt{While} statements, in order to enable more sophisticated static and dynamic analyses such as symbolic execution.
Another limitation, related to LiViola, stems from invisibility of
where a certain key for a receive lies inside the message payload.
That can happen when the key location is only determined later during
runtime, by users specifying a \emph{wildcard} location in the harness.
If LV can not ascertain keys of different receives may conflict, it
conservatively interleaves them not to hinder coverage.
That, in turn, can compound rather quickly leading to the classic
state space explosion problem but upper bounded by IRed's state space.

The solution to the first limitation that relates to conditionals is to add a field that holds an Abstract Syntax Tree (AST), generated by
scalameta~\cite{scala:meta} quasi-quotes, of the functional style conditions.
That is for some analyses such as symbolic execution to be able to determine the satisfiability of certain path conditions~\footnote{A path condition is the set of control-flow statement's conditions, e.g. an if or while statements, when satisfied (or not), the scheduler can force the execution (or not to execute) of code blocks in the program along that path, i.e. their code blocks, across said statements.}.

One solution to the second limitation is to provide a function to be executed by LV each time to determine the location of targeted keys/variables based on information available during runtime.
That, however, will require some, not many, modifications to LiViola to enable such feature.

%

%




\subsection{Discussion}

In this section we discuss few aspects regarding when our algorithms would be most effective at catching bugs, a comparison to the modern commonly used technique of concolic execution\footnote{Symbolic execution paired with runtime partial runs to fulfill unknown information during static time but that are known only during dynamic/runtime time.}, and we end the discussion by some static and dynamic frameworks that can be integrated with ours to achieve more analyses.

We rely on the classification of bugs mentioned in~\cite{bugs-in-actors} to comment the effectiveness of our technique at catching said bugs.
Lopez et. al.~\cite{bugs-in-actors} describe the following classes of actor-specific bugs that, in turn, are sub categorized further:
\begin{itemize}
  \item \textbf{Lack of Progress:} which includes communication deadlocks (two or more actors are blocked forever waiting for each other to do something), behavioral deadlocks (two or more actors waiting for a message to make progress, i.e. a message is never sent to allow for progress), live locks (similar to a deadlock but while involved actor's states are changing but without making progress).
  \item \textbf{Message Protocol Violation:} message order violation (where two or more actors exchange messages in a way violating the intended protocol), bad message interleaving (happens when a message is processed between the processing of two messages that are intended to be processed one after another), memory inconsistency (when different actors have different view of a conceptually shared resource).
\end{itemize}
Our algorithms address, and catches bugs most effectively, in the second class i.e. message protocol violations.
On the other hand, our algorithms will exhibit communication deadlocks, behavioral deadlocks, and/or live locks, in the case of lack of progress class of bugs.
Assuming we have implemented algorithms for catching the first class' bugs, we would use these algorithms to first detect and remove that class of bugs. Only then, when the implementation is free from said bugs, we would use algorithms presented in this work.
Our framework does not limit the ability from developing algorithms targeting lack of progress class of bugs.
It, actually, has facilities to ease that task, by making it explicit which statements do block and whether it is blocking on external communication (e.g. a blocking get on a future, or a timed blocking statement on a future), and these blocking statements and constructs (a future) allow for the entire gradient of synchrony and asynchrony in communication and behavioral deadlocks detection.
The snapshotting feature, of the entire global state of the distributed system, of our framework allows for detecting the lack of progress in the case of a live lock.

Symbolic execution~\cite{symbolic,gklee}, and concolic testing~\cite{concolic,symbolic-agha} would be a natural extension of our work, as mentioned in \S\ref{sec:limitations}.
It is a more advanced technique that incorporates Satisfiability Modulo Theory (SMT) solvers~\cite{z3,sat4j,cvc5,smt} (or theorem provers) to determine a model that strictly follows a certain execution path.
The set of conditions that forces that path is called a \emph{path condition}.
The conditions of control flow statements, such as an \texttt{if} or a \texttt{while} statement, are formulated as an SMT query to find a model to force that execution path in the code.
This technique is more effective and more scalable than any of the DPOR methods used in this work.
It provides more coverage, better performance, and more precision at targeting a certain criteria or code blocks.
It enables more checks to be done on more involved implementations more effectively.
Some of the bugs to target with concolic execution includes, but not limited to, dead code detection, deadlock detection, data races, termination, among other things.
In other words, concolic execution is unmistakingly superior to all techniques presented in this paper.
For example, instead of enumerating many equivalent schedules as in some of the algorithms in this work, a concolic tester can systematically target mostly unique schedules without the need to enumerate many of them.
Once symbolic execution is enabled in our framework, it would be quiet interesting to implement a concolic algorithm to empirically compare it (both performance, precision, scalability) to our algorithms.
That being said, while our technique only reports real bugs, symbolic execution might report false alarms and can be spurious, i.e. report the same/related bug many times.
That is, when a bug falls on the start of a path condition, then all branches starting with the prefix of that path condition would lead to report that same bug, as many times.
For concolic testing, this can be reduced/eliminated by synthesizing a representative test of that specific bug, running the test, and making sure if the bug actually manifests in the actual run.
Symbolic execution may suffer the limitations of an SMT in case the path condition is too long having many conditions, conditions along the path have non-linear arithmetic, the path conditions involves cryptographic functions, or simply the path conditions involves some computation outside of an SMT solver's capabilities.
Still, more than one SMT solver can be mixed to complement each other's strengths.
In case some of the said limitations are still a limiting factor of symbolic
executions, they can be overcome with concolic testing.
The parts that are easy for an SMT solver (or a theorem prover) to figure out would still be solved symbolically; along with inputs that lead to that part of the code and that was determined by solving the constraints of that path condition(s).
The parts that limits symbolic execution are to be test-ran (the concrete part of concolic) by the concolic tester, and hence the name concolic for CONCrete and symbOLIC.
That enables the concolic tester to overcome the limitations of the symbolic execution and make progress towards corner cases that are hard to hit otherwise.
Hitting the hard corner cases is a weakness for all algorithms of this paper as
the input to the systems' implementations (and through the scheduler) is done by
the user, rather than a symbolic executor.

%
%
Hence, we made sure that only the very advanced analyses may need to call into external static and/or dynamic frameworks to analyze the model or augment the operation of algorithms/schedulers.
The model composes actors/agents from a local state (a mapping construct with garbage collection that models the variables to values mapping in the scope), an activation frame stack (for function calls statement types with each activation frame having a \texttt{LocalState} object), and a mapping from message received to actions to be performed.
There are other features but to simplify the overview for the reader, we omit those.
The action is nothing but nested sequences of statement types.
Each statement type wraps an actual statement in the form of a function to call and contains all the meta data a static/dynamic analyzing algorithm needs to access.
All of the above is extensible, i.e. if more meta data were needed for a certain advanced algorithm the user definitely can extend those and add more meta data (i.e. fields/attributes to that class).
Analyzing nested lists of statements, we think, is simple enough not to involve any other framework for the majority of tasks but still can be if the user chooses to.
Similarly, algorithms and their auxiliary methods are all defined and ready to be used\footnote{We highly encourage the reader to read the operational semantics of
DS2~\cite{ds2-opsem} to realize and appreciate the capabilities there in. A simpler model paper for an overview is in~\cite{ds2-model}}.
However, we understand for example that a more advanced user (or formal methods expert) would want to use e.g. SMTs that have Java/Scala APIs in their schedules/algorithms.
That is possible, and there is nothing to interfere with their desire to involve static and/or dynamic analysis frameworks/tools.
An example we tried before is to diagram the message flow from between actors using Graphviz~\cite{graphviz} via Graph4s~\cite{graph4s}, similar to what was mentioned as MFG (message flow graphs) in Shian Li et al work~\cite{bse}.
Similarly, we visualized the states for a sample run of several of the algorithms, e.g. WGL, while we were at the debugging stages of these algorithms.
We took a look at the Backwards Symbolic Execution (BSE) by Shian Li et al~\cite{bse} and we do not see any thing preventing the integration of such an advanced form of symbolic execution.
The snapshot feature of DS2 definitely enables going back and forth in time without limits while exploring the subject distributed systems written in its model (a simple and extensible domain specific sanguage - DSL - as mentioned earlier).
As a matter of fact, multiple more complex tools can be integrated into one
scheduler and they can hand-off saved states to each other to do different
exploration techniques in one algorithm; hence, forming a collection of the
strengths of each collaboratively.
That is, the algorithm can save the state of the system, using DS2's lightweight snapshot feature, then BSE can do some kind of symbolic execution and after that returning the results.
At this point, another snapshot can be taken, then either of these snapshots given to another algorithm to do another kind of analysis on the system, taking another snapshot after done and reporting back and so on.
There is literally no limit but the host physical limits (e.g. amount of memory) on which that system is analyzed.
After all, BSE, MFG, and DS2 all define formal operational semantics by which they can be more intimately integrated.
%

\section{Related Work}
\label{sec:related-work}
In this section we taxonomize several related works. Covering both
linearizability, model checking, tracing, monitoring, verification, proofs, \dots
etc.

\noindent\textbf{Linearizability.}
Linearizability has long been presented and extensively studied in
literature; from the decision procedure perspective (checking) in the
original paper by Herlihy~\cite{herlihy:linearizability}, Jeanette Wing
and Gong~\cite{wg:linearizability}, and in WG
Lowe~\cite{lowe:wgl:linearizability}.
%
%
Testing for Linearizability~\cite{testing:for:linearizability}
developed and evaluated \emph{five} algorithms for \emph{randomly
testing} concurrent data types for linearizability violations, four of
which are new. Also, four of which are generic while one is specific
to concurrent queues.
%
Winter et al.~\cite{observational:models}
derive an approach based on the instruction re-ordering rules for
weak-memory models hardware in order to enable the re-use of existing
methods and tool support for linearizability checking. Specifically,
they target programs running on TSO
(Total Store Order) and XC (Relaxed Consistency - cache coherence
model) weak memory models.
%
Ozkan et al.~\cite{hitting:families} show that linearizability of a
history is often witnessed/realized by a schedule that re-orders small
number of operations ($< 5$), prioritizing schedules with fewer
operations first in the exploration. The algorithm characterizes
families of schedules of certain length of operations (depth) that is
guaranteed to cover all linearizability witnesses of that depth.
In our work, we experimented with minimal invocations that
allow the subject systems to exhibit all externally observable
behaviors, namely two and three invocations.
In recent years, Kyle Kingsbury the author of Jepsen~\cite{jepsen}, came up with
a new library, called Elle~\cite{elle}, to analyze histories produced by the former in order to find
consistency violations.
It is a sound framework, but since some information may be missing from observed
histories, not complete.
Hadzilacos et al~\cite{linearizable-reg-not-terminating} invalidate a conjuncture that says if we replace an atomic object in an algorithm by another object that is linearizable, then the algorithm stays the same.
One result of their work is that in randomized algorithms, when an atomic
register was replaced by a linearizable one, it lead to violating the
all-important property of termination with probability of 1.
Hence, they propose a new stronger type of register linearizability called write strong-linearizability.
It is strictly stronger than (plain) linearizability but strictly weaker than
strong linearizability and it fixes the above.
Bashari et al~\cite{linearizable-snapshot} states that in most algorithms n-processes updating different array locations in an array, a scan would produce a linearizable snapshot of the array.
However, that requires a $O(n)$ scan operation of the array.
They came up with an approach to produce such array in a constant time
complexity, and a $O(\mathtxt{log}\ n)$ observe and update operations, hence
improving the performance.
Sela et al~\cite{linearizability-typo} point out and provide an amendment to the original linearizability paper.
The typo addresses the issue of handling invocations in volatile memory setups and hence it was significant.

\noindent\textbf{Model Checking.}
Doolan et al.~\cite{spin:scalability:improve:auto:linearizability}
studied the SPIN~\cite{spin} model checker algorithm to understand the
scalability issues in an effort to scale \emph{automatic
linearizability checking} and without manual
specification of linearization points by the users. They also provide
proof-of-concept implementation. Our work does that,
also, and without manual specification of linearization points due to
the atomic nature of the actor-model actions in our model. However,
our algorithms are generic and can be applied to other problems that
can be mapped to race conditions checking and without the need to
write a separate model e.g. in Promela~\cite{promela} (the
implementation is the model).
%
Vechev et al.~\cite{ibm:experience:model:check:linearizability:with:spin}
provide an experience report summarizing first experience with
model checking linearizability. It was the first work to achieve that
with non-fixed linearization points.
%
Zhang et al.~\cite{scalable:auto:linearizability:checking} employed model
checking, partial order reduction, and symmetry between threads to
reduce the state space to model check for linearizability.
SAMC!~\cite{samc} is a model checking tool targeting message-reordering,
crashes of processes, and reboots deep bugs in distributed systems.
It requires semantic annotation to reduce the systematic exploration of
state space.
Our tool-chain is similar in the sense that it uses the operational
semantics to reduce the search space but without the need for users to
manually enter annotations.
Chong et al~\cite{model-checking-aws} describes a style of applying symbolic
model checking developed over the course of four years at Amazon Web Services
(AWS), lists lessons learned, and provides a list of proofs developed throughout
their experience developing for Amazon's AWS.

\noindent\textbf{Tracing.}
{\c{C}}irisci er al.~\cite{root:causing:linearizability:violation}
propose an approach that points the root-cause of linearizability
violations in the form of code blocks whose atomicity is required to
restore linearizability. That is, the problem can be reduced to
identifying minimal root causes of conflict serializability violation
in an error trace combined with a heuristic to find out which is the
more likely cause of the linearizability violation.
%
Zhang et al.~\cite{localization:linearizability:coarse} present a
tool called CGVT to build a small test case that is sufficient enough
for reproducing a linearizability fault. Based on a possibly long
history that was deemed non-linearizable, the tool locates the
offending operations and synthesizes a minimal test-case for further
investigation.
%
Zhang et al.~\cite{localization:linearizability:fine} provide a better
approach than CGTV by coming up with what is called
\emph{critical data race sequence} (CDRS) that side steps the
shortcoming of the coarse-grained interleaving when linearizability is
violated. The new fine-grained trace model helps in better localizing
the linearizability violations using labeled-tree model of program
executions. They implement and evaluate their approach in another
(subsequent to CGTV) tool called FGVT (Fine-Grained-Veri-Trace).

\noindent\textbf{Quasi Linearizability.}
Zhang et al.~\cite{round:up} on runtime checking for
\emph{quasi linearizability}. This is a more relaxed form of
linearizability. The authors of a tool called Inspect implemented a
fully automatic approach using LLVM to detect and report \emph{real}
violation of quasi linearizability.
%
Adhikari et al.~\cite{verifying:quasi:linearizability} verify
quantitative relaxation of \emph{quasi linearizability} of an
implementation \emph{model} of the data structure. It is based on
checking the refinement relation between the implementation and a
specification model. They implement and evaluate their approach in a
framework called PAT verification framework.

\noindent\textbf{Fixing Linearizability.}
Liu et al.~\cite{flint:fixing:linearizability} address the problem of
fixing non-linearizable \emph{composed operations} such that they
behave atomically in concurrent data structures. The algorithm (Flint)
accepts a non-linearizable composed-operations on a map. Its output,
if it succeeds at fixing the operations, is a linearizable composed
operation that is equivalent to a sequential data structure
execution. The effectiveness of the algorithm is 96\% based on 48
incorrect input compositions.

\noindent\textbf{Verification.}
Liang et al.~\cite{modular:veri:linearizability:non:fixed:points} propose a
program logic with a lightweight instrumentation mechanism which can
verify algorithms with non-fixed linearization points. This work was
evaluated on various classic algorithms some of which used in
\texttt{java.util.concurrent} package.
Bouajjani et al.~\cite{concurrent:priority:queues:linearizability}
consider concurrent priority queues, fundamental to many
multi-threaded applications such as task scheduling. It shows that
verifying linearizability of such implementations can be reduced to
control-state reachability. This result makes verifying said
data-structures in the context of unbounded number of threads
decidable.

\noindent\textbf{Runtime Monitoring.}
Emmi et al.~\cite{michael:emmi:2017} leverage an
observation about properties that admit polynomial-time (instead of
exponential time) linearizability monitoring for certain concurrent
collections data types, e.g. queues, sets, stacks, and maps. It uses
these properties to reduce linearizability to Horn
satisfiability. This work is the first in linearizability
\emph{monitoring} that is sound, complete, and tractable.
Emmi et al.~\cite{michael:emmi:2020} identify an optimization for
weak-consistency checking that relies on a \emph{minimal} visibility
relations that adhere to various constraints of the given
criteria. Hence, saving time instead of exponential enumerations of
possible visibility relations among the linearized operations. This,
as the work before it, is a monitoring approach.

\noindent\textbf{Proofs/Proving.}
Henzinger et al.~\cite{aspect:oriented:linearizability:proofs} argue
that the non-monolithic approaches based on linearization points
(automatic or manual) are both complicated and do not scale well
e.g. in optimistic updates. The work proposes a more modular
alternative approach of checking linearizability of concurrent queue
algorithms. Hence, reducing linearizability proofs of concurrent
queues to four basic properties, each of which can be proven
independently by simpler arguments.
%
Sergey et al.~\cite{hoare:style:linearizability:specification} propose a uniform
alternative to other approaches in the form of Hoare logic, which can explicitly
capture the interference of threads in an
auxiliary state. This work implements the mechanized proof methodology
in a Coq-based tool and verifies some implementations with
non-standard conditions of concurrency-aware linearizability,
quiescent and quantitative quiescent consistency.

\noindent\textbf{Compositionality of Linearizability.}
P-compositionality statically decomposes concurrent objects' (ADTs')
histories based on different operations' target keys into sub-histories
(one per key) to check for
linearizability~\cite{lulian:linearizability}.
LiViola, our algorithm, was inspired by the latter but we applied this
concept at the more critical scheduler level, as we found out later in
the experiments.
\noindent\textbf{Scheduling.}
Recently, an approach that uses invariants and dynamic scheduling to
achieve consistency-aware scheduling for weakly consistent
geo-replicated data stores has been
presented~\cite{maryam:dynamic:causally:consistent:approach},
evaluated, and found effective.
It was implemented and evaluated in a model checker for Antidote
DB~\cite{antidote:db}.

\noindent\textbf{Causality.}
Prior to that there was what is called Causal
Linearizability~\cite{causal:linearizability}.
Given some constraints on the clients, a more generic causality
checking can be achieved.
The actor model (also our model) conforms to such constraint on a
per-process (per actor) scheduled task.
Maximum Causality Reduction~\cite{mcr:maximum:causality:reduction}
uses causality information on the trace level to reduce the state
space to explore for TSO (Total Store Order) and PSO (Partial Store
Order) to check for Sequential Consistency.
Bita~\cite{bita} is an algorithm used to generate causally-consistent
schedules of receives.

\noindent\textbf{Checking Histories for Linearizability Violations.}
As for the WGL algorithm implementations, there are several other than
ours.
A Go-implementation~\cite{linearizability:checker:go:porcupine} that
requires invariants input by users to the tool, and another CPP
implementation~\cite{linearizability:checker:cpp} that works on
already output traces exist.
A similar tool-chain to ours, called
Jepsen~\cite{kyle:blog,jepsen,jepsen:repo} (generates
histories/traces) and its engine Knossos~\cite{jepsen:knossos}
(generates schedules) is widely used to check key-value stores
linearizability, black box fuzzing.
It needs significant amount of work from the user studying how the
subject system works and then devises a test harness, which takes
months of insightful work.
It handles faults while our algorithms do not; although, they can if
extended due to the model being readily prepared for simulating
faults.
LineUp~\cite{lineup}, Microsoft Research's primary C\#/.NET concurrent
objects tester, existed for a long time and it is effective at
exposing concurrent data structures linearizability violations.

\noindent\textbf{Fault Injection.}
Peter Alvaro's work on lineage driven fault
injection~\cite{alvaro:lineage:driven:fault:injection}, consistency
without borders~\cite{alvaro:consistency:borders}, Automating Failure
Testing Research At Internet
Scale~\cite{alvaro:failure:testing:internet:scale} was an
inspiration in designing and steering our framework and its extensibility.

\noindent\textbf{Runtime Scheduling.}
ARTful~\cite{artful-schedulers} is a model for user-defined schedulers
targeting various high performance computing (HPC) systems.
HPC systems, a specialized kind of distributed systems, usually experience some
load balancing issues and this work allows users to regain control over wasted
resources.
Experiments in this work has been conducted on OpenMP~\cite{openmp} and
Charm++~\cite{charm} runtime systems.

\noindent\textbf{Specification Languages.}
Pluscal~\cite{pluscal} is an algorithm language that is used to
specify distributed systems.
Users need to hand write models in this language to enable the runtime
to check desired properties.

\noindent\textbf{Learning Based Approaches.} Mukherjee et
al~\cite{ml-concurrency-testing} developed a technique for controlled
concurrency testing (CCT) using machine learning to address scale of the state
space in concurrent programs interleavings. This work developed a framework,
called QL, where they rely on Q-Learning algorithm to decide the next action to
be explored. That, in turn, is influenced by the previously selected actions in
the exploration. This work was benchmarked against a set of microbenchmarks,
complex protocols, and production cloud services and performed well. In future
works, we may employ similar techniques, i.e. machine learning, to compare
against such algorithms. Another work, Marc\'en et
al~\cite{ml-model-driven-engineering} tackled the integration problem for
machine learning classifiers to augment model driven development (MDD) in two
real industrial setups and it paid off. Our framework still does not have the
synthesis functionality. However, we plan to explore this direction in future
work since it showed encouraging results in many other setups e.g. github
co-pilot as well.


\section{Concluding Remarks}
\label{sec:conclusion}
In this work, we used only four out of sixteen semantics rules our
model provides, namely: Schedule, Consume, Send, and
Message-Reordering.
%
%
One can extend these algorithms and utilize them to address more
problems.
%
In the near future, we would like to add more benchmarks, schedulers
and address more systems faults.
%
In addition, we want to remove all model limitations to enable further
static and/or dynamic introspection.
Also, all our expectations we theorized were met and exceeded.
However, there are still room to improve IRed and LiViola.
Some heuristics that may yield these improvements include the following:
1) prioritizing longer causal-chains execution earlier in the
exploration tree, and 2) introspection into code to produce longer
causal chains that involve as many enabled receives as possible, and
3) the conflicting keys (in LiViola) can be viewed as program
variables and hence it extends to other than key-value stores.
%
%
%
After conducting this experiment, we realized that many
concurrency bugs can reduce/mapped to race condition detection,
e.g. LiViola and linearizability.
%
Last, and foremost, we have learned the following lessons out of this
work:
\begin{enumerate}
\item \textbf{Lesson 1: } This work informed the fact that focusing on
  schedules pruning is more effective than focusing on improving
  linearizability checking itself.

\item \textbf{Lesson 2:} Any approach that simply tries to maximize
  ``coverage'' in the spaces of schedules or histories is not likely to
  yield bugs unless it is efficient enough to provide near-full coverage

\item \textbf{Lesson 3:} Chronological uniqueness of histories can
  mislead that worse algorithms perform better (being too repetitive
  amplifies the perception of finding more bugs by revealing the same
  bug/history more repetitively).

\item \textbf{Lesson 4:} Layering complexity cleanly enables
  addressing more of it, more easily, and modularily. This is exemplified
  by using only 4 operational semantics rules, and future work adding
  more complexities using more semantics rules in future specialized
  schedulers.
\end{enumerate}

\noindent\textbf{Future Work.} At the end, we would like to mention some of the
potential future direction(s) for us. We find that implementing some of the
termination detection algorithms by Dan Plyukhin et
al~\cite{termination-scalable,termination-decentralized} interesting to show and
test more of the capabilities of our framework.

\noindent\textbf{Acknowledgments.} We would like to thank Zvonimir Rakamaric
for steering the work towards this fruitful direction, and the
following who contributed significantly to the project: Heath French,
Zepeng Zhao, Jarkko Savela, Anushree Singh.

\noindent\textbf{NSF Funding.} This work has been partially funded by
both NSF grant CCF 1302449, CCF 1956106 and NSF grant CSR 1421726.



\bibliography{liviola}

\begin{thebibliography}{100}
\providecommand \doibase [0]{http://dx.doi.org/}%

\bibitem{akka}
{Akka}. \url{http://akka.io/}, Retrieved May 3, 2016.

\bibitem{erlang}
{Erlang Programming Language}. 2018.
\newblock \url{https://www.erlang.org/}.

\bibitem{orleans-paper}
Bernstein PA, Bykov S, Geller A, Kliot G, Thelin J. Orleans: Distributed
  Virtual Actors for Programmability and Scalability. Tech. Rep.
  MSR-TR-2014-41,  2014.

\bibitem{actor-model-partisan}
Meiklejohn CS, Miller H, Alvaro P. {PARTISAN}: Scaling the Distributed Actor
  Runtime. In: {USENIX} Association; 2019; Renton, WA\string: 63--76.

\bibitem{akka-fortnite}
{Inside Fortnite’s Massive Data Analytics Pipeline}.
  \url{https://www.datanami.com/2018/07/31/inside-fortnites-massive-data-analytics-pipeline/},
  Retrieved May 5, 2020.

\bibitem{orleans-halo}
{Who is using Orleans}.
  \url{https://dotnet.github.io/orleans/Community/Who-Is-Using-Orleans.html},
  Retrieved May 5, 2020.

\bibitem{bita}
Tasharofi S, Pradel M, Lin Y, Johnson R. Bita: Coverage-guided, automatic
  testing of actor programs. In: ; 2013\string: 114-124

\bibitem{agha:transdpor}
Tasharofi S, Karmani RK, Lauterburg S, Legay A, Marinov D, Agha G. {TransDPOR}:
  A Novel Dynamic Partial-order Reduction Technique for Testing Actor Programs.
  In: FMOODS'12/FORTE'12. Springer-Verlag; 2012; Berlin, Heidelberg\string:
  219--234

\bibitem{basset}
Lauterburg S, Karmani RK, Marinov D, Agha G. Basset: a tool for systematic
  testing of actor programs. In:  Roman G, Hoek v.~dA. \kern-2pt, eds. {\it
  Proceedings of the 18th {ACM} {SIGSOFT} International Symposium on
  Foundations of Software Engineering, 2010, Santa Fe, NM, USA, November 7-11,
  2010}{ACM}; 2010\string: 363--364

\bibitem{verdi}
Wilcox JR, Woos D, Panchekha P, et al. Verdi: A Framework for Implementing and
  Formally Verifying Distributed Systems. {\it ACM SIGPLAN Notices}
  2015\string; 50(6)\string: 357--368.
\newblock \href {\doibase 10.1145/2813885.2737958} {doi:
  10.1145/2813885.2737958}

\bibitem{coq}
Coq, Inria. \url{https://coq.inria.fr/}, Retrieved Feb 27, 2016.

\bibitem{jepsen}
{Distributed Systems Safety Research}. \url{https://jepsen.io/}, Retrieved Feb
  27, 2018.

\bibitem{herlihy:linearizability}
Herlihy MP, Wing JM. Linearizability: A Correctness Condition for Concurrent
  Objects. {\it ACM Trans. Program. Lang. Syst.} 1990\string; 12(3)\string:
  463--492.
\newblock \href {\doibase 10.1145/78969.78972} {doi: 10.1145/78969.78972}

\bibitem{pcomp}
Horn A, Kroening D. Faster linearizability checking via
  {\textdollar}P{\textdollar}-compositionality. {\it CoRR} 2015\string;
  abs/1504.00204.

\bibitem{wg:linearizability}
Wing J, Gong C. Testing and Verifying Concurrent Objects. {\it Journal of
  Parallel and Distributed Computing} 1993\string; 17(1)\string: 164 - 182.
\newblock \href {\doibase https://doi.org/10.1006/jpdc.1993.1015} {doi:
  https://doi.org/10.1006/jpdc.1993.1015}

\bibitem{raft}
Ongaro D, Ousterhout J. In Search of an Understandable Consensus Algorithm. In:
  USENIX ATC'14. USENIX Association; 2014; Berkeley, CA, USA\string: 305--320.

\bibitem{viewstamped.revisited}
Liskov B, Cowling J. Viewstamped replication revisited.  2012.

\bibitem{paxos-simple}
Lamport L. {Paxos Made Simple}. {\it SIGACT News} 2001\string; 32(4)\string:
  51--58.

\bibitem{lulian:linearizability}
Singh V, Neamtiu I, Gupta R. Proving Concurrent Data Structures Linearizable.
  In: ; 2016\string: 230-240

\bibitem{flanagan:dpor}
Flanagan C, Godefroid P. Dynamic Partial-Order Reduction for Model Checking
  Software. In: POPL ’05. Association for Computing Machinery; 2005; New
  York, NY, USA\string: 110–121

\bibitem{ds2:repo}
{DS2 - Declarative Specification of Distributed Systems}.
  \url{https://gitlab.com/mohd_sm81/ds2}, Retrieved Jan 20, 2021.

\bibitem{dynamo}
DeCandia G, Hastorun D, Jampani M, et al. {D}ynamo: {A}mazon's {H}ighly
  {A}vailable {K}ey-value {S}tore. {\it SIGOPS Operating Systems Review}
  2007\string; 41(6)\string: 205--220.
\newblock \href {\doibase 10.1145/1323293.1294281} {doi:
  10.1145/1323293.1294281}

\bibitem{cassandra}
{The Apache Software Foundation} . {Apache Cassandra}.
  \url{http://cassandra.apache.org/}; .

\bibitem{giffords-algorithm}
Gifford DK. Weighted Voting for Replicated Data. In: SOSP ’79. Association
  for Computing Machinery; 1979; New York, NY, USA\string: 150–162

\bibitem{benchmark:openchord:heath}
{OpenChord Akka Implementation, in Java and Scala}.
  \url{https://bitbucket.org/onedash/akka-open-chord}, Retrieved April 8, 2018.

\bibitem{benchmark:openchord:zepeng}
{OpenChord Akka Implementation, Scala only}.
  \url{https://github.com/allenfromu/Open-Chord-Scala}, Retrieved April 8,
  2018.

\bibitem{make-chord-correct}
Zave P. How to Make Chord Correct (Using a Stable Base). {\it CoRR}
  2015\string; abs/1502.06461.

\bibitem{chord}
Stoica I, Morris R, Karger D, Kaashoek MF, Balakrishnan H. Chord: A Scalable
  Peer-to-peer Lookup Service for Internet Applications. In: SIGCOMM '01. ACM;
  2001; New York, NY, USA\string: 149--160

\bibitem{chandra:paxos-live}
Chandra TD, Griesemer R, Redstone J. {Paxos Made Live: An Engineering
  Perspective}. In: PODC '07. ACM; 2007; New York, NY, USA\string: 398--407

\bibitem{paxos}
Lamport L. The Part-time Parliament. {\it ACM Transactions on Computer Systems}
  1998\string; 16(2).

\bibitem{benchmark:paxos}
{Paxos Akka Implementation}.
  \url{https://github.com/allenfromu/Multi-Paxos-UDP}, Retrieved April 8, 2018.

\bibitem{ds2-opsem}
Al-Mahfoudh MS, Gopalakrishnan G, Stutsman R. Operational Semantics for the
  Rigorous Analysis of Distributed Systems. In:  Rubin SH, Bouabana-Tebibel T.
  \kern-2pt, eds. {\it Quality Software Through Reuse and Integration}Springer
  International Publishing; 2018; Cham\string: 209--231.

\bibitem{ds2-model}
Al-Mahfoudh M, Gopalakrishnan G, Stutsman R. Toward Rigorous Design of
  Domain-specific Distributed Systems. In: FormaliSE '16. ACM; 2016; New York,
  NY, USA\string: 42--48

\bibitem{ds2-and-opsem-report}
{DS2} Official Website. 2016.
\newblock \url{http://formalverification.cs.utah.edu/ds2}, Retrieved Jan 31,
  2016.

\bibitem{agha:actors}
Agha G. {\it {Actors: A Model of Concurrent Computation in Distributed
  Systems}}. PhD thesis. MIT,  1985.

\bibitem{hewitt:actors}
Hewitt C, Bishop P, Steiger R. A Universal Modular ACTOR Formalism for
  Artificial Intelligence. In: IJCAI'73. Morgan Kaufmann Publishers Inc.; 1973;
  San Francisco, CA, USA\string: 235--245.

\bibitem{actor-model-wiki}
{Actor Model - Wikipedia}. \url{https://en.wikipedia.org/wiki/Actor_model},
  Retrieved May 13, 2020.

\bibitem{samc}
Leesatapornwongsa T, Hao M, Joshi P, Lukman JF, Gunawi HS. {SAMC}:
  Semantic-aware Model Checking for Fast Discovery of Deep Bugs in Cloud
  Systems. In: OSDI'14. USENIX Association; 2014; Berkeley, CA, USA\string:
  399--414.

\bibitem{pavlo2011}
Pavlo A, Jones EP, Zdonik S. On Predictive Modeling for Optimizing Transaction
  Execution in Parallel {OLTP} Systems. {\it PVLDB} 2011\string; 5\string:
  85--96.

\bibitem{mc}
Jhala R, Majumdar R. Software Model Checking. {\it ACM Computing Surveys}
  2009\string; 41(4)\string: 21:1--21:54.
\newblock \href {\doibase 10.1145/1592434.1592438} {doi:
  10.1145/1592434.1592438}

\bibitem{modist}
Yang J, Chen T, Wu M, et al. {MODIST:} Transparent Model Checking of Unmodified
  Distributed Systems. In: {USENIX} Association; 2009\string: 213--228.

\bibitem{10.1145/1047659.1040315}
Flanagan C, Godefroid P. Dynamic Partial-Order Reduction for Model Checking
  Software. {\it SIGPLAN Not.} 2005\string; 40(1)\string: 110–121.
\newblock \href {\doibase 10.1145/1047659.1040315} {doi:
  10.1145/1047659.1040315}

\bibitem{rt-actors}
Nielsen B, Agha G. Semantics for an Actor-Based Real-Time Language. In:  {Naval
  Surface Warfare Center Dahlgren Division/IEEE (ed.)} . \kern-2pt, ed. {\it
  Proceedings of the 4th International Workshop on Parallel and Distributed
  Real-Time Systems (WPDRTS), Honolulu, Hawaii, April 1995}<Forlag uden navn>;
  1995\string: 223--228.
\newblock Semantics for an Actor-Based Real-Time Language ; Conference date:
  19-05-2010.

\bibitem{rebecca}
{Rebeca Modeling Language}. \url{https://rebeca-lang.org/}, Retrieved Jun 9,
  2022.

\bibitem{lowe:wgl:linearizability}
Lowe G. Testing for linearizability. {\it Concurrency and Computation: Practice
  and Experience}\string; 29(4)\string: e3928.
\newblock \href {\doibase 10.1002/cpe.3928} {doi: 10.1002/cpe.3928}

\bibitem{delay-bounded}
Emmi M, Qadeer S, Rakamaric Z. Delay-Bounded Scheduling. Tech. Rep.
  MSR-TR-2010-123,  2010.

\bibitem{linearizability:checker:cpp}
{Linearizability Checker}.
  \url{https://github.com/ahorn/linearizability-checker}, Retrieved April 10,
  2018.

\bibitem{viewstamped.replication.vs.others}
Van~Renesse R, Schiper N, Schneider FB. Vive la diff{\'e}rence: Paxos vs.
  viewstamped replication vs. zab. {\it IEEE Transactions on Dependable and
  Secure Computing} 2014\string; 12(4)\string: 472--484.

\bibitem{zab}
Junqueira FP, Reed BC, Serafini M. Zab: High-performance broadcast for
  primary-backup systems. In: ; 2011\string: 245-256

\bibitem{scala:meta}
{Scalameta - Library to read, analyze, transform and generate Scala programs}.
  \url{https://scalameta.org/}, Retrieved Jan 20, 2021.

\bibitem{bugs-in-actors}
López C, Marr S, Mössenböck H, Gonzalez~Boix E. A Study of Concurrency Bugs
  and Advanced Development Support for Actor-based Programs.  2017.

\bibitem{symbolic}
{Symbolic Execution}. \url{https://en.wikipedia.org/wiki/Symbolic_execution},
  Retrieved Jun 19, 2022.

\bibitem{gklee}
Li P, Li G, Gopalakrishnan G. Practical Symbolic Race Checking of GPU Programs.
  .

\bibitem{concolic}
{Concolic Testing}. \url{https://en.wikipedia.org/wiki/Concolic_testing},
  Retrieved Jun 19, 2022.

\bibitem{symbolic-agha}
Sen K, Agha G. Automated Systematic Testing of Open Distributed Programs. In:
  Baresi L, Heckel R. \kern-2pt, eds. {\it Fundamental Approaches to Software
  Engineering}Springer Berlin Heidelberg; 2006; Berlin, Heidelberg\string:
  339--356.

\bibitem{z3}
{Z3 SMT Sovler}. \url{https://github.com/Z3Prover/z3}, Retrieved April 10,
  2018.

\bibitem{sat4j}
{SAT4J}. \url{http://www.sat4j.org/}, Retrieved April 10, 2018.

\bibitem{cvc5}
{CVC5}. \url{https://cvc5.github.io/}, Retrieved Jun 20, 2022.

\bibitem{smt}
{Satisfiability modulo theories}. \url{https://github.com/Z3Prover/z3},
  Retrieved Jun 20, 2022.

\bibitem{graphviz}
{Graphviz}. \url{https://graphviz.org/}, Retrieved Jun 22, 2022.

\bibitem{graph4s}
Graph for Scala. Retrieved Jan 31, 2016.

\bibitem{bse}
Li S, Hariri F, Agha G. Targeted test generation for actor systems. In:
  Millstein T. \kern-2pt, ed. {\it 32nd European Conference on Object-Oriented
  Programming, ECOOP 2018}Leibniz International Proceedings in Informatics,
  LIPIcs. Schloss Dagstuhl- Leibniz-Zentrum fur Informatik GmbH, Dagstuhl
  Publishing; 2018

\bibitem{testing:for:linearizability}
Lowe G. Testing for linearizability. {\it Concurrency and Computation: Practice
  and Experience} 2017\string; 29(4)\string: e3928.
\newblock e3928 cpe.3928\href {\doibase https://doi.org/10.1002/cpe.3928} {doi:
  https://doi.org/10.1002/cpe.3928}

\bibitem{observational:models}
{Winter} K, {Smith} G, {Derrick} J. Observational Models for Linearizability
  Checking on Weak Memory Models. In: ; 2018\string: 100-107

\bibitem{hitting:families}
Ozkan BK, Majumdar R, Niksic F. Checking Linearizability Using Hitting
  Families. In: PPoPP '19. Association for Computing Machinery; 2019; New York,
  NY, USA\string: 366–377

\bibitem{elle}
Kingsbury K. Elle: Finding Isolation Violations in Real-World Databases. In:
  PODC'21. Association for Computing Machinery; 2021; New York, NY, USA\string:
  7

\bibitem{linearizable-reg-not-terminating}
Hadzilacos V, Hu X, Toueg S. On Register Linearizability and Termination. In:
  PODC'21. Association for Computing Machinery; 2021; New York, NY, USA\string:
  521–531

\bibitem{linearizable-snapshot}
Bashari B, Woelfel P. An Efficient Adaptive Partial Snapshot Implementation.
  In: PODC'21. Association for Computing Machinery; 2021; New York, NY,
  USA\string: 545–555

\bibitem{linearizability-typo}
Sela G, Herlihy M, Petrank E. Brief Announcement: Linearizability: A Typo. In:
  PODC'21. Association for Computing Machinery; 2021; New York, NY, USA\string:
  561–564

\bibitem{spin:scalability:improve:auto:linearizability}
Doolan P, Smith G, Zhang C, Krishnan P. Improving the Scalability of Automatic
  Linearizability Checking in SPIN. In:  Duan Z, Ong L. \kern-2pt, eds. {\it
  Formal Methods and Software Engineering}Springer International Publishing;
  2017; Cham\string: 105--121.

\bibitem{spin}
Holzmann GJ. Logic Verification of {ANSI-C} Code with {SPIN}. In: . 1885 of
  {\it Lecture Notes in Computer Science}. Springer; 2000\string: 131--147.

\bibitem{promela}
{Spin}. \url{http://spinroot.com/spin/whatispin.html}, Retrieved Feb 27, 2016.

\bibitem{ibm:experience:model:check:linearizability:with:spin}
Vechev M, Yahav E, Yorsh G. Experience with Model Checking Linearizability. In:
   P{\u{a}}s{\u{a}}reanu CS. \kern-2pt, ed. {\it Model Checking
  Software}Springer Berlin Heidelberg; 2009; Berlin, Heidelberg\string:
  261--278.

\bibitem{scalable:auto:linearizability:checking}
Zhang S. Scalable automatic linearizability checking. In: IEEE Computer
  Society; 2011; Los Alamitos, CA, USA\string: 1185-1187

\bibitem{model-checking-aws}
Chong N, Cook B, Eidelman J, et al. Code-level model checking in the software
  development workflow at Amazon Web Services. {\it Software: Practice and
  Experience} 2021\string; 51(4)\string: 772-797.
\newblock \href {\doibase https://doi.org/10.1002/spe.2949} {doi:
  https://doi.org/10.1002/spe.2949}

\bibitem{root:causing:linearizability:violation}
{\c{C}}irisci B, Enea C, Farzan A, Mutluergil SO. Root Causing Linearizability
  Violations. In:  Lahiri SK, Wang C. \kern-2pt, eds. {\it Computer Aided
  Verification}Springer International Publishing; 2020; Cham\string: 350--375.

\bibitem{localization:linearizability:coarse}
Zhang Z, Wu P, Zhang Y. Localization of Linearizability Faults on the
  Coarse-Grained Level. {\it International Journal of Software Engineering and
  Knowledge Engineering} 2017\string; 27(09n10)\string: 1483-1505.
\newblock \href {\doibase 10.1142/S0218194017400071} {doi:
  10.1142/S0218194017400071}

\bibitem{localization:linearizability:fine}
Chen Y, Zhang Z, Wu P, Zhang Y. Interleaving-Tree Based Fine-Grained
  Linearizability Fault Localization. In:  Feng X, M{\"u}ller-Olm M, Yang Z.
  \kern-2pt, eds. {\it Dependable Software Engineering. Theories, Tools, and
  Applications}Springer International Publishing; 2018; Cham\string: 108--126.

\bibitem{round:up}
{Lu Zhang} , {Chattopadhyay} A, {Wang} C. Round-up: Runtime checking quasi
  linearizability of concurrent data structures. In: ; 2013\string: 4-14

\bibitem{verifying:quasi:linearizability}
Adhikari K, Street J, Wang C, Liu Y, Zhang S. Verifying a Quantitative
  Relaxation of Linearizability via Refinement. In:  Bartocci E, Ramakrishnan
  CR. \kern-2pt, eds. {\it Model Checking Software}Springer Berlin Heidelberg;
  2013; Berlin, Heidelberg\string: 24--42.

\bibitem{flint:fixing:linearizability}
Liu P, Tripp O, Zhang X. Flint: Fixing Linearizability Violations. {\it SIGPLAN
  Not.} 2014\string; 49(10)\string: 543–560.
\newblock \href {\doibase 10.1145/2714064.2660217} {doi:
  10.1145/2714064.2660217}

\bibitem{modular:veri:linearizability:non:fixed:points}
Liang H, Feng X. Modular Verification of Linearizability with Non-Fixed
  Linearization Points. {\it SIGPLAN Not.} 2013\string; 48(6)\string:
  459–470.
\newblock \href {\doibase 10.1145/2499370.2462189} {doi:
  10.1145/2499370.2462189}

\bibitem{concurrent:priority:queues:linearizability}
Bouajjani A, Enea C, Wang C. Checking Linearizability of Concurrent Priority
  Queues. {\it CoRR} 2017\string; abs/1707.00639.

\bibitem{michael:emmi:2017}
Emmi M, Enea C. Sound, Complete, and Tractable Linearizability Monitoring for
  Concurrent Collections. {\it Proc. ACM Program. Lang.} 2017\string; 2(POPL).
\newblock \href {\doibase 10.1145/3158113} {doi: 10.1145/3158113}

\bibitem{michael:emmi:2020}
Emmi M, Enea C. Monitoring Weak Consistency. In:  Chockler H, Weissenbacher G.
  \kern-2pt, eds. {\it Computer Aided Verification}Springer International
  Publishing; 2018; Cham\string: 487--506.

\bibitem{aspect:oriented:linearizability:proofs}
Henzinger TA, Sezgin A, Vafeiadis V. Aspect-Oriented Linearizability Proofs.
  In:  D'Argenio PR, Melgratti H. \kern-2pt, eds. {\it CONCUR 2013 --
  Concurrency Theory}Springer Berlin Heidelberg; 2013; Berlin,
  Heidelberg\string: 242--256.

\bibitem{hoare:style:linearizability:specification}
Sergey I, Nanevski A, Banerjee A, Delbianco GA. Hoare-Style Specifications as
  Correctness Conditions for Non-Linearizable Concurrent Objects. In: OOPSLA
  2016. Association for Computing Machinery; 2016; New York, NY, USA\string:
  92–110

\bibitem{maryam:dynamic:causally:consistent:approach}
Dabaghchian M, Rakamaric Z, Ozkan BK, Mutlu E, Tasiran S. Consistency-Aware
  Scheduling for Weakly Consistent Programs. {\it SIGSOFT Softw. Eng. Notes}
  2018\string; 42(4)\string: 1–5.
\newblock \href {\doibase 10.1145/3149485.3149493} {doi:
  10.1145/3149485.3149493}

\bibitem{antidote:db}
Shapiro M, Bieniusa A, Preguiça N, Balegas V, Meiklejohn C. Just-Right
  Consistency: reconciling availability and safety. 2018.

\bibitem{causal:linearizability}
Doherty S, Derrick J. Linearizability and Causality. In:  De~Nicola R, K{\"u}hn
  E. \kern-2pt, eds. {\it Software Engineering and Formal Methods}Springer
  International Publishing; 2016; Cham\string: 45--60.

\bibitem{mcr:maximum:causality:reduction}
Huang S, Huang J. Maximal Causality Reduction for TSO and PSO. {\it SIGPLAN
  Not.} 2016\string; 51(10)\string: 447--461.
\newblock \href {\doibase 10.1145/3022671.2984025} {doi:
  10.1145/3022671.2984025}

\bibitem{linearizability:checker:go:porcupine}
{Porcupine}. \url{https://github.com/anishathalye/porcupine}, Retrieved April
  10, 2018.

\bibitem{kyle:blog}
{Jepsen}. \url{https://aphyr.com/tags/jepsen}, Retrieved April 7, 2018.

\bibitem{jepsen:repo}
{Jepsen}. \url{https://github.com/jepsen-io/jepsen}, Retrieved April 10, 2018.

\bibitem{jepsen:knossos}
{Knossos: Verifies the linearizability of experimentally accessible
  histories.}. \url{https://github.com/jepsen-io/knossos}, Retrieved Apr 17,
  2018.

\bibitem{lineup}
Burckhardt C, Musuvathi M, Tan R. Line-up: A Complete and Automatic
  Linearizability Checker. {\it SIGPLAN Not.} 2010\string; 45(6)\string:
  330--340.
\newblock \href {\doibase 10.1145/1809028.1806634} {doi:
  10.1145/1809028.1806634}

\bibitem{alvaro:lineage:driven:fault:injection}
Alvaro P, Rosen J, Hellerstein JM. Lineage-driven Fault Injection. In: SIGMOD
  '15. ACM; 2015; New York, NY, USA\string: 331--346

\bibitem{alvaro:consistency:borders}
Alvaro P, Bailis P, Conway N, Hellerstein JM. Consistency Without Borders. In:
  SOCC '13. ACM; 2013; New York, NY, USA\string: 23:1--23:10

\bibitem{alvaro:failure:testing:internet:scale}
Alvaro P, Andrus K, Sanden C, Rosenthal C, Basiri A, Hochstein L. Automating
  Failure Testing Research at Internet Scale. In: SoCC '16. ACM; 2016; New
  York, NY, USA\string: 17--28

\bibitem{artful-schedulers}
Santana A, Freitas V, Castro M, Pilla LL, Méhaut JF. ARTful: A model for
  user-defined schedulers targeting multiple high-performance computing runtime
  systems. {\it Software: Practice and Experience} 2021\string; 51(7)\string:
  1622-1638.
\newblock \href {\doibase https://doi.org/10.1002/spe.2977} {doi:
  https://doi.org/10.1002/spe.2977}

\bibitem{openmp}
{OpenMP}. \url{https://www.openmp.org/}, Retrieved Jun 22, 2022.

\bibitem{charm}
{Charm++}. \url{https://charmplusplus.org/}, Retrieved Jun 22, 2022.

\bibitem{pluscal}
Lamport L. The PlusCal Algorithm Language. {\it Theoretical Aspects of
  Computing-ICTAC 2009, Martin Leucker and Carroll Morgan editors. Lecture
  Notes in Computer Science, number 5684, 36-60.} 2009.

\bibitem{ml-concurrency-testing}
Mukherjee S, Deligiannis P, Biswas A, Lal A. Learning-Based Controlled
  Concurrency Testing. {\it Proc. ACM Program. Lang.} 2020\string; 4(OOPSLA).
\newblock \href {\doibase 10.1145/3428298} {doi: 10.1145/3428298}

\bibitem{ml-model-driven-engineering}
Marcén AC, Pérez F, Pastor O, Cetina C. Evaluating the benefits of empowering
  model-driven development with a machine learning classifier. {\it Software:
  Practice and Experience} 2022\string; 52(11)\string: 2439-2455.
\newblock \href {\doibase https://doi.org/10.1002/spe.3133} {doi:
  https://doi.org/10.1002/spe.3133}

\bibitem{termination-scalable}
Plyukhin D, Agha G. Scalable Termination Detection for Distributed Actor
  Systems. In:  Konnov I, Kov{\'{a}}cs L. \kern-2pt, eds. {\it 31st
  International Conference on Concurrency Theory, {CONCUR} 2020, September 1-4,
  2020, Vienna, Austria (Virtual Conference)}. 171 of {\it LIPIcs}. Schloss
  Dagstuhl - Leibniz-Zentrum f{\"{u}}r Informatik; 2020\string: 11:1--11:23

\bibitem{termination-decentralized}
Plyukhin D, Agha G. A Scalable Algorithm for Decentralized Actor Termination
  Detection. {\it Log. Methods Comput. Sci.} 2022\string; 18(1).
\newblock \href {\doibase 10.46298/lmcs-18(1:39)2022} {doi:
  10.46298/lmcs-18(1:39)2022}

\end{thebibliography}

\appendix
\section{Raw Data for the Benchmarks}
\label{benchmarks:appendix}
\begin{table}[b!]
  \centering
  \scriptsize
  \begin{tabular}{| l | l |}
    \hline
    \textbf{Symbol} & \textbf{Meaning} \\\hline\hline
    \textbf{EX}   &  \makecell{Exhaustive Scheduler} \\ \hline
    \textbf{SR}   &  \makecell{Systematic Random Scheduler} \\ \hline
    \textbf{DB}   &  \makecell{Delay-Bounded Scheduler} \\ \hline
    \textbf{DP}   &  \makecell{DPOR - Dynamic Partial Order Reducing Scheduler} \\ \hline
    \textbf{TD}   &  \makecell{TransDPOR - Transitive DPOR} \\ \hline
    \textbf{IR}   &  \makecell{IRed Scheduler - generic precise causality tracking scheduler} \\ \hline
    \textbf{LV}   &  \makecell{LiViola - specialized Linearizability Violation Scheduler, and transitive race scheduler} \\ \hline
    \textbf{\#Agents}& \makecell{Number of agents in the system} \\ \hline
    \textbf{\#Rtry}  & \makecell{Number of retries per request} \\ \hline
    \textbf{Qrm}  & \makecell{Number of agents forming to form a quorum in the system} \\ \hline
    \textbf{\#Inv}& \makecell{Number of invocations, 2R+1W means 2 reads and 1 write} \\ \hline
    \textbf{\#S}  & \makecell{Number of schedules}\\ \hline
    \textbf{\#IH} & \makecell{Number of incomplete histories} \\ \hline
    \textbf{\#UH} & \makecell{Number of \emph{chronologically} unique histories} \\ \hline
    \textbf{\#NL} & \makecell{Number of Non-Linearizable unique histories} \\ \hline
    \textbf{NL/UH}& \makecell{Number of Non-Linearizable unique histories to total number of unique histories ratio, or the \emph{quality}} \\ \hline
    \textbf{UH/S} & \makecell{Number of unique histories to the number of schedules ratio, or the \emph{progression}} \\ \hline
    \textbf{NL/S} & \makecell{Number of Non-Linearizable unique histories to the number of  schedules ratio, or the \emph{precision}}\\ \hline
    \textbf{TS}   & \makecell{The time spent by the scheduler to generate all schedules} \\ \hline
    \textbf{ST}   & \makecell{The approximate time if the scheduler is to produce the same schedules but statelessly} \\ \hline
    \textbf{TC}   & \makecell{The time spent to check all unique histories in the configuration} \\ \hline
    \textbf{TT}   & \makecell{The total time spent to both generate schedules and check all unique histories, $TT = TS + TC$} \\ \hline
    \textbf{\#HF} & \makecell{Number of histories before catching the first buggy (non-linearizable) history} \\ \hline
    \textbf{TF}   & \makecell{The time to hit the first buggy (non-linearizable) history} \\ \hline
  \end{tabular}
  \caption{The legend of all symbols used in the results tables}
  \label{tab:legend}
\end{table}


\begin{table}[htb!]
  \centering
  \scriptsize
  \begin{tabular}{| m{2cm} | c || c  c  c  c  c  c  c ||} \cline{2-9}
    \multicolumn{1}{l|}{}& \textbf{\#Inv.}
    & \multicolumn{7}{c |}{\textbf{2R+1W}}\\ \cline{2-9}
    \multicolumn{1}{l|}{}& \textbf{Alg.}
    & \textbf{EX} & \textbf{SR} & \textbf{DB} & \textbf{DP} & \textbf{TD} & \textbf{IR} & \textbf{LV} \\ \hline
    \multirow{13}{2cm}{\textbf{Distributed Register}
    \newline \mbox{\#Agents = 2}
    \newline \mbox{\#Rtry = 3}
    \newline \mbox{Qrm = 2} }   
                         & \textbf{\#S}  & 4,200  & 100K   & 4,200  & 2,984   & 2,906   & 2,906   & 88    \\ \cline{2-9}
                         & \textbf{\#IH} & 4,195  &  0     & 4,195  & 2,979   & 2,901   & 2,901   & 83    \\ \cline{2-9}
                         & \textbf{\#UH} &   37   &  26    &   37   &   37    &   37    &   37    & 22    \\ \cline{2-9}
                         & \textbf{\#NL} &   0    &  0     &   0    &   0     &   0     &   0     &  0    \\ \cline{2-9}
                         & \textbf{NL/UH}&  0.0 \%&  0.0\% &   0.0\%&    0.0\%&  0.0\%  &   0.0\% & 0.0\% \\ \cline{2-9}
                         & \textbf{UH/S} &  0.88\%&  0.03\%&  0.88\%&  1.24\% &  1.27\% &  1.27\% & 25.0\%\\ \cline{2-9}
                         & \textbf{NL/S} & 0.0\%  &  0.0\% & 0.0\%  &   0.0\% &  0.0\%  &  0.0\%  & 0.0\% \\ \cline{2-9}
                         & \textbf{TS}   & 0:06   & 4:48   & 0:06   & 0:05    & 0:05    & 0:04    & 0:00  \\ \cline{2-9}
                         & \textbf{ST}   & 0:18   & 1:52   & 0:21   & 0:12    & 0:13    & 0:17    & 0:00  \\ \cline{2-9}
                         & \textbf{TC}   & 0:00   & 0:00   & 0:00   & 0:00    & 0:00    & 0:00    & 0:00  \\ \cline{2-9}
                         & \textbf{TT}   & 0:06   & 4:48   & 0:06   & 0:05    & 0:05    & 0:04    & 0:00  \\ \cline{2-9}
                         & \textbf{\#HF} &  -     &  -     & -      & -       &  -      &   -     &  -    \\ \cline{2-9}
                         & \textbf{TF}   & 0:00   & 0:00   & 0:00   & 0:00    & 0:00    & 0:00    & 0:00  \\ \hline
  \end{tabular}
  \caption{Correct distributed register results for the 3-receives harness}
  \label{tab:dr1}
\end{table}


\begin{table}[htb!]
  \centering
  \scriptsize
  \begin{tabular}{| m{2cm} | c || c  c  c  c  c  c  c ||} \cline{2-9}
    \multicolumn{1}{l|}{}& \textbf{\#Inv.}
    & \multicolumn{7}{c |}{\textbf{2R+2W}}\\ \cline{2-9}
    \multicolumn{1}{l|}{}& \textbf{Alg.}
    & \textbf{EX} & \textbf{SR} & \textbf{DB} & \textbf{DP} & \textbf{TD} & \textbf{IR} & \textbf{LV} \\ \hline
    \multirow{13}{2cm}{\textbf{Distributed Register}
    \newline \mbox{\#Agents = 2}
    \newline \mbox{\#Rtry = 3}
    \newline \mbox{Qrm = 2} }   
                                  & \textbf{\#S}  & 50K    & 100K  & 50K   &  50K & 50K   &  50K  & 7,236 \\ \cline{2-9}
                                  & \textbf{\#IH} & 50K    & 100K  & 50K   & 50K  & 50K   &  50K  & 7,236 \\ \cline{2-9}
                                  & \textbf{\#UH} & 8      &   91  &  8    &  8   &  8    &  8    &  147  \\ \cline{2-9}
                                  & \textbf{\#NL} & 0      &  0    &  0    &  0   &  0    &  0    &  0    \\ \cline{2-9}
                                  & \textbf{NL/UH}&  0.0\% & 0.0\% & 0.0\% & 0.0\%& 0.0\% & 0.0\% & 0.0\% \\ \cline{2-9}
                                  & \textbf{UH/S} & 0.02\% & 0.09\%& 0.02\%&0.02\%& 0.02\%& 0.02\%& 2.03\%\\ \cline{2-9}
                                  & \textbf{NL/S} &  0.0\% & 0.0\% & 0.0\% &0.0\% & 0.0\% & 0.0\% & 0.0\% \\ \cline{2-9}
                                  & \textbf{TS}   & 1:27   & 5:49  & 1:22  & 1:36 & 1:34  & 1:34  & 0:09  \\ \cline{2-9}
                                  & \textbf{ST}   & 9:27   & 2:55  & 6:27  & 7:09 & 7:42  & 7:05  & 0:34  \\ \cline{2-9}
                                  & \textbf{TC}   & 0:00   & 0:00  & 0:00  & 0:00 & 0:00  & 0:00  & 0:00  \\ \cline{2-9}
                                  & \textbf{TT}   & 1:27   & 5:49  & 1:22  & 1:36 & 1:34  & 1:34  & 0:09  \\ \cline{2-9}
                                  & \textbf{\#HF} & -      &  -    &  -    &  -   &  -    &  -    &  -    \\ \cline{2-9}
                                  & \textbf{TF}   & 0:00   & 0:00  & 0:00  & 0:00 & 0:00  & 0:00  & 0:00  \\ \cline{2-9} \hline
  \end{tabular}
  \caption{Correct distributed register results for the 4-receives harness}
  \label{tab:dr2}
\end{table}


\begin{table}[htb!]
  \centering
  \scriptsize
  \begin{tabular}{| m{2cm} | c || c  c  c  c  c  c  c ||} \cline{2-9}
    \multicolumn{1}{l|}{}& \textbf{\#Inv.}
    & \multicolumn{7}{c |}{\textbf{2R+1W}}\\ \cline{2-9}
    \multicolumn{1}{l|}{}& \textbf{Alg.}
    & \textbf{EX} & \textbf{SR} & \textbf{DB} & \textbf{DP} & \textbf{TD} & \textbf{IR} & \textbf{LV} \\ \hline
    \multirow{13}{2cm}{\textbf{Erroneous Distributed Register}
    \newline \mbox{\#Agents = 3}
    \newline \mbox{\#Rtry = 3}
    \newline \mbox{Qrm = 2} }   
                         & \textbf{\#S}  &  252   &  100K  &  252   &  20     &   20    &  14     &  14    \\ \cline{2-9}
                         & \textbf{\#IH} &  247   &   0    &  247   &  15     &   15    &  11     &  11    \\ \cline{2-9}
                         & \textbf{\#UH} &  11    &   66   &  11    &  11     &   12    &  8      &  8     \\ \cline{2-9}
                         & \textbf{\#NL} &   2    &   0    &   2    &   3     &   4     &  2      &  2     \\ \cline{2-9}
                         & \textbf{NL/UH}& 18.18\%&  0.0\% & 18.18\%&  27.27\%& 33.33\% &  25.0\% & 25.0\% \\ \cline{2-9}
                         & \textbf{UH/S} & 4.37\% &  0.07\%& 4.37\% &  55.0\% & 60.0\%  &  57.14\%&57.14\% \\ \cline{2-9}
                         & \textbf{NL/S} & 0.79\% &  0.0\% & 0.79\% &  15.0\% & 20.0\%  &  14.29\%&14.29\% \\ \cline{2-9}
                         & \textbf{TS}   & 0:00   & 4:09   & 0:00   & 0:00    & 0:00    & 0:00    &  0:00  \\ \cline{2-9}
                         & \textbf{ST}   & 0:01   & 1:39   & 0:01   & 0:00    & 0:00    & 0:00    &  0:00  \\ \cline{2-9}
                         & \textbf{TC}   & 0:00   & 0:00   & 0:00   & 0:00    & 0:00    & 0:00    &  0:00  \\ \cline{2-9}
                         & \textbf{TT}   & 0:00   & 4:09   & 0:00   & 0:00    & 0:00    & 0:00    &  0:00  \\ \cline{2-9}
                         & \textbf{\#HF} &  2     &  -     &  2     &   0     &  0      &   1     &  1     \\ \cline{2-9}
                         & \textbf{TF}   & 0:00   & 0:00   & 0:00   & 0:00    & 0:00    & 0:00    & 0:00   \\ \cline{2-9} \hline
  \end{tabular}
  \caption{Erroneous distributed register results for the 3-receives harness}
  \label{tab:edr1}
\end{table}


\begin{table}[htb!]
  \centering
  \scriptsize
  \begin{tabular}{| m{2cm} | c || c  c  c  c  c  c  c ||} \cline{2-9}
    \multicolumn{1}{l|}{}& \textbf{\#Inv.}
    & \multicolumn{7}{c |}{\textbf{2R+2W}}\\ \cline{2-9}
    \multicolumn{1}{l|}{}& \textbf{Alg.}
    & \textbf{EX} & \textbf{SR} & \textbf{DB} & \textbf{DP} & \textbf{TD} & \textbf{IR} & \textbf{LV} \\ \hline
    \multirow{13}{2cm}{\textbf{Erroneous Distributed Register}
    \newline \mbox{\#Agents = 3}
    \newline \mbox{\#Rtry = 3}
    \newline \mbox{Qrm = 2} }   
                         & \textbf{\#S}  &  50K   &  100K &  50K  & 50K   &  50K  &30,063 & 28,517\\ \cline{2-9}
                         & \textbf{\#IH} & 49,996 &   0   & 49,996&49,894 & 49,875&29,995 & 28,441\\ \cline{2-9}
                         & \textbf{\#UH} &  20    &  459  &  18   &  79   &  152  &  97   &  116  \\ \cline{2-9}
                         & \textbf{\#NL} &  0     &  0    &  0    &  11   &  18   & 6     &  10   \\ \cline{2-9}
                         & \textbf{NL/UH}&  0.0\% & 0.0\% & 0.0\% &13.92\%&11.84\%& 6.19\%& 8.62\%\\ \cline{2-9}
                         & \textbf{UH/S} & 0.04\% & 0.46\%& 0.04\%& 0.16\%& 0.3\% & 0.32\%& 0.41\%\\ \cline{2-9}
                         & \textbf{NL/S} & 0.0\%  & 0.0\% & 0.0\% & 0.02\%& 0.04\%& 0.02\%& 0.04\%\\ \cline{2-9}
                         & \textbf{TS}   & 1:31   & 4:34  & 1:50  & 1:48  & 1:46  & 1:11  & 1:13  \\ \cline{2-9}
                         & \textbf{ST}   & 5:19   & 2:00  & 5:56  & 6:29  & 5:46  & 3:01  & 3:02  \\ \cline{2-9}
                         & \textbf{TC}   & 0:00   & 0:00  & 0:00  & 0:00  & 0:00  & 0:00  & 0:00  \\ \cline{2-9}
                         & \textbf{TT}   & o1:31  & 4:34  & 1:50  & 1:48  & 1:46  & 1:11  & 1:13  \\ \cline{2-9}
                         & \textbf{\#HF} &  -     &  -    &   -   &  0    &  0    &  0    &  0    \\ \cline{2-9}
                         & \textbf{TF}   & 0:00   & 0:00  & 0:00  & 0:00  & 0:00  & 0:00  & 0:00  \\ \cline{2-9} \hline
  \end{tabular}
  \caption{Erroneous distributed register results for the 4-receives harness}
  \label{tab:edr2}
\end{table}


\begin{table}[htb!]
  \centering
  \scriptsize
  \begin{tabular}{| m{2cm} | c || c  c  c  c  c  c  c ||} \cline{2-9}
    \multicolumn{1}{l|}{}& \textbf{\#Inv.}
    & \multicolumn{7}{c |}{\textbf{2R+1W}}\\ \cline{2-9}
    \multicolumn{1}{l|}{}& \textbf{Alg.}
    & \textbf{EX} & \textbf{SR} & \textbf{DB} & \textbf{DP} & \textbf{TD} & \textbf{IR} & \textbf{LV} \\ \hline
    \multirow{13}{2cm}{\textbf{Another Distributed Register}
    \newline \mbox{\#Agents = 2}
    \newline \mbox{\#Rtry = 3}
    \newline \mbox{Qrm = 2} }   
                         & \textbf{\#S}  & 1,680  & 100K   & 1,680  & 908     &  818    &  802    & 802    \\ \cline{2-9}
                         & \textbf{\#IH} & 887    &  0     &   887  & 462     &  419    &  412    & 412    \\ \cline{2-9}
                         & \textbf{\#UH} & 148    &  28    &   148  & 140     &  132    &   124   & 124    \\ \cline{2-9}
                         & \textbf{\#NL} &  0     &  0     &   0    &   0     &   0     &   0     &  0     \\ \cline{2-9}
                         & \textbf{NL/UH}& 0.0\%  &  0.0\% &  0.0\% &  0.0\%  &  0.0\%  &  0.0\%  & 0.0\%  \\ \cline{2-9}
                         & \textbf{UH/S} & 8.81\% &  0.03\%& 8.81\% & 15.42\% & 16.14\% & 15.46\% &15.46\% \\ \cline{2-9}
                         & \textbf{NL/S} & 0.0\%  &  0.0\% & 0.0\%  &  0.0\%  &   0.0\% &  0.0\%  & 0.0\%  \\ \cline{2-9}
                         & \textbf{TS}   & 0:22   & 8:54   & 0:20   & 0:06    & 0:07    & 0:07    & 0:07   \\ \cline{2-9}
                         & \textbf{ST}   & 1:12   & 1:10   & 1:07   & 0:19    & 0:21    & 0:20    & 0:22   \\ \cline{2-9}
                         & \textbf{TC}   & 0:00   & 0:00   & 0:00   & 0:00    & 0:00    & 0:00    & 0:00   \\ \cline{2-9}
                         & \textbf{TT}   & 0:22   & 8:54   & 0:20   & 0:06    & 0:07    & 0:07    & 0:07   \\ \cline{2-9}
                         & \textbf{\#HF} &   -    &  -     & -      & -       & -       & -       & -      \\ \cline{2-9}
                         & \textbf{TF}   & 0:00   & 0:00   & 0:00   & 0:00    & 0:00    & 0:00    & 0:00   \\ \cline{2-9} \hline
  \end{tabular}
  \caption{Another Correct distributed register results for the 3-receives harness}
  \label{tab:adr1}
\end{table}


\begin{table}[htb!]
  \centering
  \scriptsize
  \begin{tabular}{| m{2cm} | c || c  c  c  c  c  c  c ||} \cline{2-9}
    \multicolumn{1}{l|}{}& \textbf{\#Inv.}
    & \multicolumn{7}{c |}{\textbf{2R+2W}}\\ \cline{2-9}
    \multicolumn{1}{l|}{}& \textbf{Alg.}
    & \textbf{EX} & \textbf{SR} & \textbf{DB} & \textbf{DP} & \textbf{TD} & \textbf{IR} & \textbf{LV} \\ \hline
    \multirow{13}{2cm}{\textbf{Another Distributed Register}
    \newline \mbox{\#Agents = 2}
    \newline \mbox{\#Rtry = 3}
    \newline \mbox{Qrm = 2} }   
                         & \textbf{\#S}  &  50K   &  100K &  50K  &  50K   & 50K   &  50K  & 50K   \\ \cline{2-9}
                         & \textbf{\#IH} & 42,890 & 95,060& 42,890& 40,152 & 39,564& 39,884& 39,884\\ \cline{2-9}
                         & \textbf{\#UH} &   718  &  131  &  718  &  1,610 & 1,876 & 1,907 & 1,907 \\ \cline{2-9}
                         & \textbf{\#NL} &    0   &  0    &  0    &  0     & 0     &  0    & 0     \\ \cline{2-9}
                         & \textbf{NL/UH}&  0.0\% & 0.0\% & 0.0\% &0.0\%   & 0.0\% & 0.0\% & 0.0\% \\ \cline{2-9}
                         & \textbf{UH/S} & 1.44\% & 0.13\%& 1.44\%&3.22\%  & 3.75\%& 3.81\%& 3.81\%\\ \cline{2-9}
                         & \textbf{NL/S} &  0.0\% & 0.0\% & 0.0\% &0.0\%   & 0.0\% & 0.0\% & 0.0\% \\ \cline{2-9}
                         & \textbf{TS}   & 693:56 & 4:44  & 660:41&876:36  & 835:39& 859:53& 845:30\\ \cline{2-9}
                         & \textbf{ST}   &2635:48 & 1:52  &2608:10&2871:33 &2703:56&2889:04&2857:07\\ \cline{2-9}
                         & \textbf{TC}   & 0:00   & 0:00  & 0:00  & 0:00   & 0:00  & 0:00  & 0:00  \\ \cline{2-9}
                         & \textbf{TT}   & 693:56 & 4:44  & 660:41& 876:36 & 835:39& 859:53& 845:30\\ \cline{2-9}
                         & \textbf{\#HF} &  -     & -     & -     &  -     &  -    &  -    &  -    \\ \cline{2-9}
                         & \textbf{TF}   & 0:00   & 0:00  & 0:00  & 0:00   & 0:00  & 0:00  & 0:00  \\ \cline{2-9} \hline
  \end{tabular}
  \caption{Another Correct distributed register results for the 4-receives harness}
  \label{tab:adr2}
\end{table}


\begin{table}[htb!]
  \centering
  \scriptsize
  \begin{tabular}{| m{2cm} | c || c  c  c  c  c  c  c ||} \cline{2-9}
    \multicolumn{1}{l|}{}& \textbf{\#Inv.}
    & \multicolumn{7}{c |}{\textbf{2R+1W}}\\ \cline{2-9}
    \multicolumn{1}{l|}{}& \textbf{Alg.}
    & \textbf{EX} & \textbf{SR} & \textbf{DB} & \textbf{DP} & \textbf{TD} & \textbf{IR} & \textbf{LV} \\ \hline
    \multirow{13}{2cm}{\textbf{Multi-Paxos}
    \newline \mbox{\#Agents = 2}
    \newline \mbox{\#Rtry = *}
    \newline \mbox{Qrm = majority} }   
                         & \textbf{\#S}  &  50K   & 100K   &  50K   &  50K    & 50K     &   50K   &  50K  \\ \cline{2-9}
                         & \textbf{\#IH} &  0     &  0     &  0     &   0     &  0      &   0     &  0    \\ \cline{2-9}
                         & \textbf{\#UH} &  14    &  26    &   14   &  12     &  12     &   12    &  12   \\ \cline{2-9}
                         & \textbf{\#NL} &   0    &  0     &   0    &   0     &   0     &   0     &   0   \\ \cline{2-9}
                         & \textbf{NL/UH}&  0.0\% &  0.0\% &  0.0\% &   0.0\% &   0.0\% &   0.0\% & 0.0\% \\ \cline{2-9}
                         & \textbf{UH/S} & 0.03\% &  0.03\%&  0.03\%&  0.02\% &   0.02\%&   0.02\%& 0.02\%\\ \cline{2-9}
                         & \textbf{NL/S} & 0.0\%  &  0.0\% &  0.0\% &  0.0\%  &   0.0\% &   0.0\% & 0,0\% \\ \cline{2-9}
                         & \textbf{TS}   & 68:19  & 6:25   & 67:51  & 74:44   & 32:15   & 80:51   & 76:13 \\ \cline{2-9}
                         & \textbf{ST}   & 306:30 & 3:22   & 318:48 & 332:29  & 149:52  &350:19   & 325:29\\ \cline{2-9}
                         & \textbf{TC}   & 0:00   & 0:00   & 0:00   & 0:00    & 0:00    & 0:00    & 0:00  \\ \cline{2-9}
                         & \textbf{TT}   & 68:19  & 6:25   & 67:51  & 74:44   & 32:15   &80:51    & 76:13 \\ \cline{2-9}
                         & \textbf{\#HF} &   -    &  -     & -      & -       & -       &   -     &  -    \\ \cline{2-9}
                         & \textbf{TF}   & 0:00   & 0:00   & 0:00   & 0:00    & 0:00    & 0:00    & 0:00  \\ \cline{2-9} \hline 
  \end{tabular}
  \caption{Multi-Paxos results for the 3-receives harness}
  \label{tab:px1}
\end{table}


\begin{table}[htb!]
  \centering
  \scriptsize
  \begin{tabular}{| m{2cm} | c || c  c  c  c  c  c  c ||} \cline{2-9}
    \multicolumn{1}{l|}{}& \textbf{\#Inv.}
    & \multicolumn{7}{c |}{\textbf{2R+2W}}\\ \cline{2-9}
    \multicolumn{1}{l|}{}& \textbf{Alg.}
    & \textbf{EX} & \textbf{SR} & \textbf{DB} & \textbf{DP} & \textbf{TD} & \textbf{IR} & \textbf{LV} \\ \hline
    \multirow{13}{2cm}{\textbf{Multi-Paxos}
    \newline \mbox{\#Agents = 2}
    \newline \mbox{\#Rtry = *}
    \newline \mbox{Qrm = majority} }   
                         & \textbf{\#S}  & 50K    &  100K & 50K   & 50K  &  50K  &  50K  &  50K  \\ \cline{2-9}
                         & \textbf{\#IH} &  0     &   0   &  0    &  0   &  0    &   0   &   0   \\ \cline{2-9}
                         & \textbf{\#UH} &  6     &  85   &  6    &  6   &  6    &   6   &   9   \\ \cline{2-9}
                         & \textbf{\#NL} &  0     &  0    &  0    &  0   &  0    &   0   &   0   \\ \cline{2-9}
                         & \textbf{NL/UH}& 0.0\%  & 0.0\% & 0.0\% &0.0\% & 0.0\% & 0.0\% & 0.0\% \\ \cline{2-9}
                         & \textbf{UH/S} & 0.01\% & 0.08\%& 0.01\%&0.01\%& 0.01\%& 0.01\%& 0.02\%\\ \cline{2-9}
                         & \textbf{NL/S} & 0.0\%  & 0.0\% & 0.0\% &0.0\% & 0.0\% & 0.0\% & 0.0\% \\ \cline{2-9}
                         & \textbf{TS}   & 35:54  & 6:12  & 35:28 & 56:50& 57:22 &70:57  & 75:48 \\ \cline{2-9}
                         & \textbf{ST}   & 286:39 & 3:40  & 307:22&581:43& 570:32&706:18 & 635:05\\ \cline{2-9}
                         & \textbf{TC}   & 0:00   & 0:00  & 0:00  & 0:00 & 0:00  & 0:00  & 0:00  \\ \cline{2-9}
                         & \textbf{TT}   & 35:54  & 6:12  & 35:28 &56:50 & 57:22 & 70:57 & 75:48 \\ \cline{2-9}
                         & \textbf{\#HF} &   -    &   -   &   -   &  -   &   -   &   -   &   -   \\ \cline{2-9}
                         & \textbf{TF}   & 0:00   & 0:00  & 0:00  & 0:00 & 0:00  & 0:00  & 0:00  \\ \cline{2-9} \hline 
  \end{tabular}
  \caption{Multi-Paxos results for the 4-receives harness}
  \label{tab:px2}
\end{table}


\begin{table}[htb!]
  \centering
  \scriptsize
  \begin{tabular}{| m{2cm} | c || c  c  c  c  c  c  c ||} \cline{2-9}
    \multicolumn{1}{l|}{}& \textbf{\#Inv.}
    & \multicolumn{7}{c |}{\textbf{2R+1W}}\\ \cline{2-9}
    \multicolumn{1}{l|}{}& \textbf{Alg.}
    & \textbf{EX} & \textbf{SR} & \textbf{DB} & \textbf{DP} & \textbf{TD} & \textbf{IR} & \textbf{LV} \\ \hline
    \multirow{13}{2cm}{\textbf{Open Chord}
    \newline \mbox{\#Agents = 2}
    \newline \mbox{\#Rtry = N/A}
    \newline \mbox{Qrm = N/A} }   
                         & \textbf{\#S}  & 18,396 &  100K  & 18,396 &  5,926  & 2,633   &  1,935  & 21    \\ \cline{2-9}
                         & \textbf{\#IH} &   0    &  0     &   0    &   0     & 0       &  0      &  0    \\ \cline{2-9}
                         & \textbf{\#UH} &  165   & 35     &  165   &   143   &  117    &  121    & 17    \\ \cline{2-9}
                         & \textbf{\#NL} &  17    &  0     &  17    &   16    &  14     &  12     &  4    \\ \cline{2-9}
                         & \textbf{NL/UH}&  10.3\%&  0.0\% & 10.3\% & 11.19\% & 11.97\% &  9.92\% &23.53\%\\ \cline{2-9}
                         & \textbf{UH/S} &  0.9\% & 0.03\% & 0.9\%  &  2.41\% & 4.44\%  &  6.25\% &80.95\%\\ \cline{2-9}
                         & \textbf{NL/S} & 0.09\% &  0.0\% & 0.09\% &  0.27\% & 0.53\%  &  0.62\% &19.05\%\\ \cline{2-9}
                         & \textbf{TS}   & 5:30   & 6:30   & 5:34   & 0:43    & 0:10    & 0:06    & 0:00  \\ \cline{2-9}
                         & \textbf{ST}   & 18:25  & 3:40   & 17:56  & 2:04    & 0:34    & 0:24    & 0:00  \\ \cline{2-9}
                         & \textbf{TC}   & 0:00   & 0:00   & 0:00   & 0:00    & 0:00    & 0:00    & 0:00  \\ \cline{2-9}
                         & \textbf{TT}   & 5:30   & 6:30   & 5:34   & 0:43    & 0:10    & 0:06    & 0:00  \\ \cline{2-9}
                         & \textbf{\#HF} &   15   &   -    &   15   &  12     &    10   &  12     &   0   \\ \cline{2-9}
                         & \textbf{TF}   & 0:00   & 0:00   & 0:00   & 0:00    & 0:00    & 0:00    & 0:00  \\ \cline{2-9} \hline
  \end{tabular}
  \caption{Open Chord results for the 3-receives harness}
  \label{tab:oc1}
\end{table}


\begin{table}[htb!]
  \centering
  \scriptsize
  \begin{tabular}{| m{2cm} | c || c  c  c  c  c  c  c ||} \cline{2-9}
    \multicolumn{1}{l|}{}& \textbf{\#Inv.}
    & \multicolumn{7}{c |}{\textbf{2R+2W}}\\ \cline{2-9}
    \multicolumn{1}{l|}{}& \textbf{Alg.}
    & \textbf{EX} & \textbf{SR} & \textbf{DB} & \textbf{DP} & \textbf{TD} & \textbf{IR} & \textbf{LV} \\ \hline
    \multirow{13}{2cm}{\textbf{Open Chord}
    \newline \mbox{\#Agents = 2}
    \newline \mbox{\#Rtry = N/A}
    \newline \mbox{Qrm = N/A} }   
                         & \textbf{\#S}  & 50K    &  100K &  50K  &  50K   &  50K  & 50K   & 71    \\ \cline{2-9}
                         & \textbf{\#IH} & 0      &   0   & 0     & 0      &  0    &  0    &  0    \\ \cline{2-9}
                         & \textbf{\#UH} & 325    & 329   & 3557  & 1073   &  1034 & 1498  &  49   \\ \cline{2-9}
                         & \textbf{\#NL} &  15    &  1    &  357  &  181   &  163  & 295   &  1    \\ \cline{2-9}
                         & \textbf{NL/UH}& 4.62\% & 0.3\% &10.04\%&16.87\% &15.76\%&19.69\%& 2.04\%\\ \cline{2-9}
                         & \textbf{UH/S} & 0.65\% & 0.33\%& 7.11\%& 2.15\% &2.07\% & 3.0\% &69.01\%\\ \cline{2-9}
                         & \textbf{NL/S} & 0.03\% & 0.0\% & 0.71\%& 0.36\% & 0.33\%& 0.59\%& 1.41\%\\ \cline{2-9}
                         & \textbf{TS}   & 73:29  & 7:40  & 34:01 & 50:48  & 60:03 & 100:43& 0:00  \\ \cline{2-9}
                         & \textbf{ST}   & 323:39 & 4:43  & 138:25& 194:59 & 214:55& 359:59& 0:00  \\ \cline{2-9}
                         & \textbf{TC}   & 0:00   & 0:00  & 0:00  & 0:00   & 0:00  & 0:00  & 0:00  \\ \cline{2-9}
                         & \textbf{TT}   & 73:29  & 7:40  & 34:02 & 50:48  & 60:03 & 100:43& 0:00  \\ \cline{2-9}
                         & \textbf{\#HF} & 15     & 168   &   1   &  2     &  9    &   1   &   22  \\ \cline{2-9}
                         & \textbf{TF}   & 0:00   & 0:00  & 0:00  & 0:00   & 0:00  & 0:00  & 0:00  \\ \cline{2-9} \hline
  \end{tabular}
  \caption{Open Chord results for the 4-receives harness}
  \label{tab:oc2}
\end{table}


\end{document}